\documentclass[aps,prb,twocolumn,nopacs,amssymb]{revtex4}

\usepackage{amsmath}

\usepackage{graphicx}
\usepackage{dcolumn}
\usepackage{bm}
\usepackage{color}

\usepackage{amsmath}
\usepackage{amssymb}

\def\be{\begin{equation}}
\def\en{\end{equation}}
\def\bea{\begin{eqnarray}}
\def\ena{\end{eqnarray}}
\def\p{\partial}
\def\ep{\epsilon}
\def\gs{\gtrsim}
\def\ls{\lesssim}

\newcommand{\av}[1]{\langle{#1}\rangle}

\newcommand{\bi}[1]{\mbox{\boldmath$#1$}}

\def\ef{\epsilon_\infty}

\begin{document}
\title{Static and dynamic theory   of  polarization under  
internal and directing  electric fields: 
Fixed-charge and fixed-potential conditions 
}  
\author{Akira Onuki\footnote{e-mail: onuki@scphys.kyoto-u.ac.jp 
(corresponding author)}}

\affiliation{
 Department of Physics, Kyoto University, Kyoto 606-8502, Japan 
}


\date{\today}

\begin{abstract} 
We present a   continuum theory on  statics and dynamics   
of polar  fluids, where    the orientational polarization ${\bi p}_1$ 
and the induced  polarization ${\bi p}_2$ are governed by 
 the Onsager directing  field ${\bi E}_d$  
and the Lorentz internal field $\bi F$, respectively. We start with a  
dielectric free energy functional $\cal F$ 
with a  cross term $\propto 
\int\hspace{-0.5mm} d{\bi r}~{\bi p}_1\cdot{\bi p}_2$, 
which was proposed by 
 Felderhof $[$J. Phys. C: Solid State Phys. {\bf 12}, 2423 
(1979)$]$.   With  this  cross-coupling, our   theory can 
 yield   the theoretical results by 
Onsager and  Kirkwood.  
 We also present  dynamic equations 
 using the functional derivatives 
$\delta {\cal F}/\delta {\bi p}_i$  
to calculate the space-time correlations of ${\bi p}_i$. 
We  then obtain analytic expressions for 
 various frequency-dependent  quantities including  the Debye formula. 
We find that the fluctuations of the total polarization 
drastically  depend on whether we fix the electrode charge  or 
the applied potential difference between  parallel metal electrodes. 
 In   the latter fixed-potential condition,  
we obtain   a nonlocal (long-range)    polarization correlation  
   inversely proportional to the cell  volume $V$, which   
 is crucial to understand   the  dielectric response. It   is   produced by  
nonlocal  charge  fluctuations on the electrode surfaces and  
  is    sensitive to the potential drops   in the Stern layers 
in small systems. These   nonlocal  correlations  in the bulk and  
on the surfaces  are closely related due to the global 
constraint of fixed potential difference. 
We also add  some  results in other boundary conditions 
including the periodic one, where nonlocal correlations 
also appear. 
\end{abstract}


\maketitle


\section{Introduction} 
The dielectric properties of polar fluids have long been studied 
 since the pioneering work by Debye\cite{Debye}, 
which is based on the  Lorentz internal field 
${\bi F}$.  Onsager\cite{Onsager} 
 introduced the directing  field\cite{Bott,Fro} ${\bi E}_d$, which  governs  
the orientational polarization ${\bi p}_1$. 
Remarkably, ${\bi E}_d$ 
   is  much smaller than $\bi F$ in magnitude for highly polar fluids, 
whereas  ${\bi E}_d={\bi F}$ in Debye's theory. In contrast, 
  the induced polarization ${\bi p}_2$ 
is proportional to  $\bi F$,  where  the proportionality 
constant is the  polarizability 
 related to the high-frequency  dielectric constant 
 $\ef$ via the 
Clausius-Mossotti relation. Accounting for the 
long-range dipolar interaction,  Onsager derived the 
so-called Onsager equation, 
which relates the dipole moment $\mu_0$ to 
$\ef$ and the static dielectric 
constant $\ep$. Soon afterwards, Kirkwood\cite{Kirk} included  
 the short-range dipole-dipole correlation in the 
dielectric response. Fr$\ddot{\rm{o}}$hlich\cite{Fro}  combined 
these  two theories  in   
the Kirkwood-Fr$\ddot{\rm{o}}$hlich (KF) equation.

Later, Felderhof\cite{Felder2} 
 presented a  free energy 
including   a cross-coupling term $\propto {\bi p}_1\cdot{\bi p}_2$, which is 
called the Lorentz term in this paper. It 
yields the Lorentz field $\bi F$  for ${\bi p}_2$ and the 
Onsager directing field ${\bi E}_d$ for ${\bi p}_1$. 
However,  many other 
authors\cite{Lee,Mat1,Li,Marcus,Kim,Tomasi,OnukiLong}   
 did not include   the Lorentz term in their free energies. 
In this study,  we derive   
   Onsager's results and the KF equation  from 
a  continuum theory based on  Felderhof's free energy by expressing 
   the polarization correlation functions  
 in simple forms.

A number of 
theories\cite{Neu,Bagchi,Wi,Elton,Ni,Mukamel,Deutch,Ful2,Gla,Cole,Zw,
Mason,Mad,Wer}   have been presented on the 
 frequency-dependent dielectric constant $\ep^*(\omega)$ at 
 frequency $\omega$. In particular, Cole\cite{Cole} 
assumed deformable molecular bonds  giving rise to atomic ${\bi p}_2$,
 while Fatuzzo and Mason\cite{Mason}  derived a frequency-dependent KF 
equation. We also mention well-developed research on the 
wave-vector-dependent  dielectric  
response\cite{Deutch,Mukamel,Ful2,Bagchi,Mad}, 
to which   the time-correlations of the Fourier components 
of ${\bi p}$  are  related. 
In dielectric measurements\cite{Elton,Fro}, $\ep^*(\omega)$ 
has been  well approximated by 
the Debye formula\cite{Debye,Fro},  
\be 
\ep^*(\omega)=\ef+ (\ep-\ef)/(1+i\omega\tau_{\rm D}),   
\en   
which involves  a single 
relaxation time  $\tau_{\rm D}$. 
However, this formula  is inadequate at  high $\omega$ 
for complex polar molecules. 
In this study, we present 
    dynamic equations for ${\bi p}_1$ and ${\bi p}_2$ 
using Felderhof's  free energy. 
They  give   three  relaxation 
times\cite{Fleming,Hubbard,Mad,Bagchi,Berne,Elton}, 
$\tau_{\rm D}$, $\tau_{\rm L}$, 
and $\tau_{\rm f}$, for  the transverse part of ${\bi p}_1$, 
the longitudinal part of ${\bi p}_1$, 
and ${\bi p}_2$, respectively, at fixed electric charges. 
Here,  $\tau_{\rm D}>\tau_{\rm L}=\tau_{\rm D}\ef/\ep\gg \tau_{\rm f}$.  Then, 
 calculating  the  time-correlation functions of ${\bi p}_i$ 
 analytically, we    obtain various dielectric 
relations including the Debye formula (1).

We should  understand  the dielectric response in the frame 
of Kubo's linear response theory\cite{Kubo,Kubo1}. He  supposed 
an externally applied time-dependent force $F(t)$ 
and its conjugate internal  variable $\cal A$ with   
the  interaction energy ${\cal H}_{\rm int}(t) =
 -{\cal A} F(t)$ in the Hamiltonian formalism. 
 In the dielectric theory, however, 
 it is not clear how to determine  $ F(t)$ and $\cal A$ 
due to the long-range electrostatic interactions 
among   the dipoles and  the electrode charges. 
 In the geometry of  parallel metal plates, 
 we can  control the surface charge $Q_0$ 
on one  electrode 
  or the applied potential difference $\Phi_a$.  
The physical consequences in these two 
conditions  are very different in statics and dynamics. 
 A number of  simulations have been 
performed at  fixed 
$\Phi_a$\cite{Le,Sprik1,T1,Takae,T2,Hau,P1,P3,P4,P5,P6,P7,Limmer,La,Sato,S1,Wang,Holm}. 
Recently, $\ep^*(\omega)$ 
 was calculated at  fixed $\Phi_a$\cite{S1}. 
Simulations have 
also been performed by 
Sprik's group\cite{Sprik,Cox} 
when the space average of the Maxwell field 
$\bi E$ or that of  the electric induction 
$\bi D$ vanishes, where the former 
 is realized in the periodic boundary condition\cite{Cai}. 
To remove 
 the  effects of the surfaces and the sample shape, 
some authors supposed 
 applied fields varying  sinusoidally in space and time 
for dielectric fluids\cite{Deutch,Mukamel}  and 
electrolytes\cite{Lebe}, where 
 the periodic boundary condition 
can be used without electrodes in 
simulation\cite{Ku,Leeuw,Lada,Lada1,Bopp}.

In this paper, we 
determine  $ F(t)$ and $\cal A$ 
 to   calculate the   relaxation functions\cite{Kubo}
  of   ${\bi p}_1$,  ${\bi p}_2$, and 
${\bi p}={\bi p}_1+{\bi p}_2$, where $ F(t)$ 
is homogeneous in space  but  can  oscillate  in time.  
We use   the  parallel plate 
geometry with  a separation length  $H$, 
where the $z$ axis is perpendicular to the 
electrode  surfaces.   
(i) First, $Q_0$  is   controlled, where   $\cal A$ is the 
integration of $p_z$ in the cell, written as $M_z^{\rm tot}$. 
In this case,  the relaxation functions 
  decay with $\tau_{\rm L}$. 
(ii) Second,  a  mesoscopic  sphere  
 in the bulk in the cell is under an {\it oscillating  directing field}, 
where $\cal A$ is the integration of 
$p_{1z}$ in the sphere.   
We then obtain    a frequency-dependent KF equation. 
(iii)  Third,  a  mesoscopic  sphere  is under 
an {\it oscillating  cavity  field}, 
 where  $\cal A$ is the integration of 
$p_{z}$ in the sphere. Then, we express 
 $\ep^*(\omega)$ in terms of  the time 
correlation of $\bi p$ 
in the long wavelength limit.  (iv) Fourth, we control 
 $\Phi_a$ with   ${\cal A}= M_z^{\rm tot}/H$. 
In this case,  the space-time correlation of $p_{z}$ 
 contains a  nonlocal (long-range) part  
 inversely proportional to the cell volume $V$, 
which   is  induced by  small fluctuations of  $Q_0$.  
Previously,  nonlocal polarization correlations  were discussed    
in different contexts\cite{Wer,Ni,Cai}. 
In fluid mixtures,   nonlocal density correlations  
$\propto V^{-1}$ generally appear 
  in the canonical\cite{Lebo}  and isobaric-isothermal\cite{OnukiP} 
ensembles. 

It is known that a significant potential drop appears   in the Stern  
layers on  solid-fluid surfaces 
\cite{Cox1,Hamann,Behrens,Hau,T2,Takae,P6,Sakuma,Maty1,Laage}. 
It gives rise to  the effective  dielectric constant,     
\be 
\ep_{\rm eff}= 4\pi Q_0 H^2/V\Phi_a=\ep/({1+ \ell_{\rm w}/H}). 
\en 
Here,   $\ell_{\rm w}$ is a surface electric 
length\cite{Cox1,Takae,Maty1}, which  is 
enlarged for $\ep\gg 1$ and is  
 of order $10$ nm for liquid water. The relation (2) holds both 
at fixed $Q_0$ and at fixed $\Phi_a$ 
and   agrees  with an experiment by Geim's group\cite{Geim}.
See a recent review on the dielectric response in  nanoconfined water 
by Mondal and Bagchi\cite{Mon}. 
We further present a frequency-dependent generalization 
of Eq.(2). We shall see that  the variance and the lifetime 
of  $M_z^{\rm tot}$  at fixed $\Phi_a$ are 
larger than those at fixed $Q_0$ 
by factors of $\ep_{\rm eff}$ and $\ep_{\rm eff}/\ef$, respectively.

We also calculate  the in-plane correlation 
of the electrode charge density $\sigma_0(x,y,t)$. 
At fixed $Q_0= \int dx dy~\sigma_0$, it has a  nonlocal term  
proportional to the inverse surface area $H/V$. 
At fixed $\Phi_a$, it additionally acquires  a  nonlocal term proportional to 
$1/V(1+\ell_{\rm w}/H)$ and equal 
 to the variance of  $M_z^{\rm tot}$. 
 which   is small but 
 greatly alters the overall dielectric response. 
These  aspects have   been overlooked  
in previous papers on the surface 
charge fluctuations\cite{Limmer,La,P7}.

The organization of this paper is as follows. 
In Sec.II, the statics of  dielectrics will be discussed, 
where  the relationship of our theory 
and  Onsager's theory will be elucidated. 
In Sec.III, 
the equal-time polarization correlations will  be calculated 
 for wave lengths longer than the molecular length and shorter than 
$H$, while  Sec.IV will present those of the 
total polarizations. In Sec.V,  
the solvation  free 
energy\cite{Lee,Li,Marcus,Kim,Tomasi}    will 
be examined in the presence of  solute charges.  
In Sec.VI, the dynamics  of ${\bi p}_1$ and ${\bi p}_2$ 
and their time-correlations will be studied. 
In Sec.VII, the dynamical linear response relations will be given. 
In Sec.VIII, the dielectric response 
and the fluctuations in  the polarizations  
and the surface charges 
will be studied at fixed $\Phi_a$. 
In Secs.IV, VI, and VIII, the fluctuations 
of the total polarizations and the nonlocal 
correlations  will be studied in other 
 boundary conditions.

\section{Polarization in linear regime}

In the continuum approach, 
we treat  a nearly incompressible, one-component 
  polar liquid   with a    
 permanent  dipole moment  $\mu_0$. 
 Electric  charges  are present 
only one the electrode surfaces 
 (which will be supposed in the fluid in Sec.V). There is  no chemical 
reaction in the fluid and on the surfaces. 
 The  net  polarization  $\bi p$ consists 
of  orientational,  atomic, 
and    electronic contributions\cite{Marcus,Lee,Felder2,Fro}. 
The  orientational one  ${\bi p}_1$  is dominant at low frequencies 
in highly polar fluids. 
The sum of the atomic and electronic   ones  
is   written as    ${\bi p}_2$ 
and is   called the  induced polarization.
  The Maxwell electric field  ${\bi E}=-\nabla\Phi$ 
is produced by  ${\bi p}= {\bi p}_1+{\bi p}_2$ and the 
electrode charges, where  $\Phi$ is  the electric potential.  
The electric induction  is 
given by  $ {\bi D}= {\bi E}+4\pi {\bi p}$ 
in the cgs units.

\subsection{Dielectric free energy }

As in Fig.1, the fluid is confined  between 
parallel metal plates  with surface  charges 
$Q_0$ and $Q_H(= -Q_0)$ at $z=0$ and $H$, respectively. 
 The lateral cell length $L$  much 
exceeds  $H$ and the edge effect is negligible. 
The average  charge density is written as 
${\bar\sigma}_0=Q_0/L^2$.  
The cell volume is  $V=L^2H$. 
In simulations,  the periodic boundary condition 
can be imposed in the $xy$ plane.  
In this section, $Q_0$ is  stationary, but  it can be 
oscillatory from  Sec.VII. 

 \begin{figure}[t]
\includegraphics[scale=1.5]{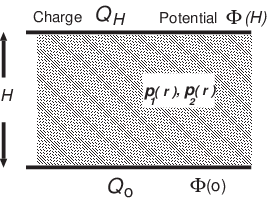}
\caption{\protect  
 Dielectric fluid with orientational polarizations ${\bi p}_1({\bi r})$ 
and induced one ${\bi p}_2({\bi r})$ between parallel metal plates, where   
$Q_0$ and  $\Phi(0)$ are the electric charge and potential, 
respectively, at $z=0$, while $Q_H(=-Q_0) $ and  $\Phi(H)$ 
 are those at $z=H$. The $z$ axis is perpendicular to the 
plates. We control $Q_0$  in the fixed-charge condition 
and $\Phi_a$ in Eq.(17) in the fixed-potential condition.}
\end{figure}

W start with  a continuum   dielectric 
free energy ${\cal F}$  bilinear in ${\bi p}_i$ and $\bi E$ 
in the linear response regime. 
It is  given by  the space  integral 
${\cal F} =\int_V d{\bi r} f $ in the cell with 
\be 
f = \frac{1}{8\pi}|{\bi E}|^2 + \frac{1}{2} a_{11}
|{\bi p}_1|^2+ a_{12}{\bi p}_1 \cdot {\bi p}_2+ 
\frac{1}{2} a_{22}|{\bi p}_2|^2,
 \en 
where $a_{ij}=a_{ji}$ are  constants   
and  the fluid-solid interaction terms 
are   not written.   The  first electrostatic term 
sensitively depends  on the   boundary condition.  
The third cross term ($\propto a_{12}$), called  the Lorentz term, 
was   introduced   by Felderhof\cite{Felder2}. 
We can further add   the gradient terms (such as 
const.$|\nabla\cdot{\bi p}|^2$)\cite{Ma,Ko,Maggs,Onukibook}  and 
 the elastic coupling terms for elastic dielectrics\cite{Eri}.

To seek equilibrium from Eq.(3), we superimpose  small 
increments $\delta {\bi E}$ and $\delta {\bi p}_i$ 
on $\bi E$ and ${\bi p}_i$ 
at fixed $a_{ij}$ 
as\cite{Eri,Maggs,Landau-e,OnukiD}   
\be 
\delta {\cal F}= \int_V\hspace{-1mm}
d{\bi r} \Big[\frac{1}{4\pi}{\bi E}\cdot\delta{\bi D}
-{\bi E}\cdot\delta{\bi p}
+ \sum_{i,j}a_{ij} {\bi p}_i\cdot \delta{\bi p}_j \Big]  ,
\en  
where $\delta{\bi D}=\delta{\bi E}+4\pi\delta{\bi p}$. 
Since ${\bi E}= -\nabla\Phi$ and $\nabla\cdot{\bi D}=0$, 
we find $\int_V d{\bi r} {\bi E}\cdot\delta{\bi D}
=0$     in the  fixed-charge condition ($\delta Q_0=0$). 
Thus, minimization of ${\cal F}$ 
 yields  
\bea 
&&\hspace{-5mm}\delta{\cal F}/\delta {\bi p}_1= 
 a_{11}{\bi p}_1+ a_{12}{\bi p}_2-{\bi E}={\bi 0}, \\
&&\hspace{-5mm}\delta{\cal F}/\delta {\bi p}_2= 
a_{22}{\bi p}_2+ a_{12}{\bi p}_1-{\bi E}={\bi 0},
\ena 
which  are solved to give ${\bi p}_i= \chi_i {\bi E}$ with 
the susceptibilities $\chi_i$ for the two polarizations. 
The net   susceptibility 
is given by  $\chi=(\ep-1)/4\pi= \chi_1+\chi_2$.    
Using the inverse  matrix $\{ a^{ij}\}$ of 
 $\{ a_{ij}\}$ we obtain  
 \bea 
&&\hspace{-5mm}
\chi_1={\sum}_{j} a^{1j}= 
 (a_{22}- a_{12})/(a_{11}a_{22}-a_{12}^2),\nonumber \\
&&\hspace{-5mm}
\chi_2= {\sum}_{j} a^{2j}  
= (a_{11}- a_{12})/(a_{11}a_{22}-a_{12}^2).
\ena  
In equilibrium   Eqs.(5) and (6) give  
the thermodynamic  dielectric free energy 
  ${\cal F}= V{\bi E}\cdot{\bi D}/8\pi$.

We introduce  the Lorentz internal field $\bi F$ by\cite{Debye}  
\be 
{\bi F}={\bi E}+ \frac{4\pi}{3} {\bi p},
\en 
which becomes    ${\bi F}= [(\ep+2)/3]{\bi E}$ in equilibrium.
Then, following Felderhof\cite{Felder2}, 
we   assume  that ${\bi p}_2$ 
is given by     
\be 
{\bi p}_2= {\bar\alpha} {\bi F}=
\frac{1}{4\pi}(\epsilon_\infty-1)
\Big( {\bi E}+ \frac{4\pi}{3} {\bi p}_1 \Big),  
\en 
where $\ef-1=4\pi{\bar\alpha}/(1-4\pi{\bar\alpha}/3)$. 
Between $\bar\alpha$ and $\ef$ we have the Clausius-Mossotti relation,    
\be 
{\bar\alpha}= n\alpha_0 = \frac{3}{4\pi}\cdot
\frac{\epsilon_\infty-1}{\epsilon_\infty+2},  
\en 
where  ${\bar\alpha}$ is 
 the  dimensionless scalar polarizability,  
   $\alpha_0$  is  the  
molecular polarizability,  and 
 $n$ is  the average dipole density.  
 For polarizable  polar fluids,  $\epsilon_\infty$ is the 
 dielectric constant at relatively high frequencies 
for which   ${\bi p}_1$ 
is negligibly small   due  to its slow relaxation (see Eq.(95)). 

We assume that Eqs.(6) and (9) both hold  
  in general situations, where  ${\bi p}_1$ can differ 
from $ \chi_1{\bi E}$.  Then,     
\be 
a_{12}=-\frac{4\pi}{3}, ~~~~
a_{22}=\frac{4\pi}{\ef-1}. 
\en 
From  Eq.(7) we can 
 express $\chi$ in terms of $a_{ij}$.  
Using    $\chi-1/a_{22}= (\ep-\ef)/4\pi$, 
   we then find   
\be 
a_{11}=\frac{4\pi}{9} \cdot \frac{\ef+2}{\epsilon-\ef}(\epsilon+2) 
-\frac{4\pi}{3} .
\en 
Now,  $\chi_1$ and $\chi_2$  are expressed  in terms of $\ep$ 
and $\ef$ as   
\be
\chi_1= \frac{3(\epsilon-\ef)}{4\pi(\ef+2)}
=\frac{3\chi}{\ef+2}-{\bar\alpha}, ~~ ~~
\chi_2=
\frac{\epsilon+2}{3} {\bar\alpha}. 
\en  
Then,   for $\ep\gg \ef$, we have 
$\chi_1\cong 3\chi/(\ef+2)$ and $\chi_2 
\cong (\ef-1)\chi/(\ef+2) 
\gg (\ef-1)/4\pi$. 
With Eqs.(11)-(13) we can also express   $f $  in a symmetrical form,   
\be 
f= \frac{1}{8\pi} |{\bi E}|^2- \frac{2\pi}{3}|{\bi p}|^2 
+  \frac{\ep+2}{6} \Big[ \frac{1}{\chi_1}|{\bi p}_1|^2+
\frac{ 1}{\chi_2}|{\bi p}_2|^2\Big] .  
 \en

We have fixed  the  surface charge 
$Q_0$ to derive  Eqs.(5) and (6). 
If the applied potential  difference $\Phi_a$ 
is fixed at a stationary value, we should 
perform the   Legendre transformation of the free energy
\cite{Landau-e,Ben,OnukiD},  
\be 
{\tilde{\cal F}}={\cal F} 
-\frac{1}{4\pi} \int_V d{\bi r} 
{\bi E}\cdot{\bi D}= {\cal F} -Q_0 \Phi_a, 
\en 
where the  surface free energy is not written  
and use is made of  $\nabla\cdot{\bi D}=0$ without ions 
in the fluid. For small changes in ${\bi p}_i$ and $\Phi_a$, 
 ${\tilde{\cal F}}$ changes  infinitesimally as 
\be 
\delta {\tilde{\cal F}}= \int_V\hspace{-1mm}
d{\bi r} \Big[-{\bi E}\cdot\delta{\bi p}
+ \sum_{i,j}a_{ij} {\bi p}_i\cdot \delta{\bi p}_j \Big]  
+ Q_0 \delta\Phi_a. 
\en  
Thus,  $\delta{\tilde{\cal F}}/\delta{\bi p}_i$ at 
fixed $\Phi_a$  are  equal to $\delta{\cal F}/\delta{\bi p}_i$ at fixed $Q_0$. 
As a result, minimization of   ${\tilde{\cal F}}$  at  fixed $\Phi_a$ 
also gives      Eqs.(5) and (6).  
In equilibrium  we have   
  ${\tilde{\cal F}}=- V{\bi E}\cdot{\bi D}/8\pi$.
 Previously,   the transformation (15) has been performed  for 
electrolytes\cite{OnukiD,Limmer,Ben}.

\subsection{Effect of surface potential drop}
 
We  remark the effect of the    potential drop 
 in the  Stern layer on a solid-fluid 
surface\cite{Hamann,Behrens,Hau,T1,T2,Takae,P6,Mon},  
where the layer    thickness $d$ is 
microscopic  and the electric field is very strong. 
The drop  at $z\cong 0$ 
consists of an  intrinsic one  $ \Phi_0^0 $ 
without electric charges and an induced one  $\sigma_0/C_0$, 
where $\sigma_0$ is the local surface charge density 
 and $C_0$ is the surface capacitance. 
The intrinsic drop arises from the molecular anisotropy 
(see Fig.1 in our previous paper\cite{T2}). 

We also  write    the surface drop in the upper Stern 
layer region $H-d<z<H$  as  
 $ -\Phi_0^H - \sigma_H/C_H$ along the $z$ axis. Then, removing 
the net  charge-free  potential drop   $\Phi_0^0-\Phi_0^H$, 
we   define the 
applied potential difference $\Phi_a$  as     
\bea 
&&\hspace{-15mm}\Phi_a= E_a H=
 {\bar\Phi}(d) -{\bar\Phi}(H-d)+ {\bar \sigma}_0/C  \nonumber\\
&&\hspace{-5mm}
=H\int_V d{\bi r}E_z({\bi r})/V+ {\bar\sigma}_0/C,
\ena
where   $1/C=1/C_0+1/C_H$ and   
${\bar\Phi}(z)$ and ${\bar\sigma}_0$ are the lateral  
averages  of  the electric potential $\Phi({\bi r})$ 
and ${\sigma}_0(x,y)$, respectively.  
In this paper, the cell integration $\int_Vd{\bi r}$ is performed 
outside the Stern layers, leading to 
the second line of  Eq.(17) (which will be a key 
relation in  Sec.VIII).

In the bulk region $d<z<H-d$, 
the electric field assumes a  bulk value  $E_b$,  where 
    $\Phi(d)-\Phi(H-d)= (H-2d)E_b$ and 
  $4\pi{{\bar\sigma}_0}=\ep E_b$. From Eq.(17) we find 
\be 
E_b  = {E_a}/({1+ \ell_{\rm w}/H}).  
\en 
Here,  $\ell_{\rm w}$ is the surface electric length
\cite{Cox1,Takae,Maty1},  
\be 
\ell_{\rm w}= \ep/(4\pi C)-2d, 
\en  
where the first term    is amplified by $\ep$ for $\ep\gg 1$. 
We then obtain the effective dielectric constant (2). 
Note that Eq.(19) can be used in the static limit. 
We will discuss  the frequency-dependent 
surface effect in Appendix A.

At  metal-water surfaces in 
the   ambient condition (at $T\cong 300$ K 
and $p\cong 1$ atm), 
 the surface capacitance  was  in a range of 5-50 $\mu$F$/$cm$^2$  
($0.45-4.5/$nm) in experiments\cite{Hamann}  
and   around  10 $\mu$F$/$cm$^2$ 
in  simulations\cite{Takae,Hau,P6}. These indicate  
 $1/4\pi C \sim 1$$~{\rm \AA}$ and  
   $\ell_{\rm w}\sim  10$ nm    for  water, 
where   $\ell_{\rm w}\cong  \chi/ C$. 
Thus, the situation 
 $d\ll H\ls \ell_{\rm w}$  can well be realized  
in     simulations\cite{Takae,Hau,P6} 
and experiments\cite{Geim}. 
See Appendix A and Sec.VIII 
for more discussions.

\subsection{Directing field and Onsager theory }

 Felderhof\cite{Felder2} introduced 
 a  coefficient $\lambda_{11}$ by  
\be 
\lambda_{11}=  1/\chi_d-a_{11}
= 1/\chi_d- (\ep+2)/3\chi_1+{4\pi}/{3},     
\en 
where  $a_{11}$ is given in Eq.(12) and 
$\chi_d$ is the   orientational susceptibility. 
 He required  the relation 
  $\delta{\cal F}/\delta{\bi p}_1= 
\chi_d^{-1} {\bi p}_1- {\bi E}_d$ using the simplest 
 form $\chi_d=n\mu_0^2/3k_BT$. From Eq.(4) we express ${\bi E}_d$ as       
\be
{\bi E}_d ={\bi E}+\lambda_{11}{\bi p}_1-a_{12} {\bi p}_2
={\bi F}-\Big( \frac{4\pi}{3}-\lambda_{11}\Big) 
{\bi p}_1, 
\en 
where   $\bi F$ is the Lorentz field (8). 
In equilibrium  we  have 
\be 
 {\bi p}_1=\chi_1 {\bi E}= \chi_d {\bi E}_d, \\
\en 
Thus, ${\bi E}_d$ is called  the  directing  field\cite{Bott} 
governing the dipole orientation.   
Debye\cite{Debye} originally   assumed  $ {\bi E}_d={\bi F}$ 
and  ${\bi p}_1= 
(n\mu_0^2/3k_BT){\bi F}$.

 Onsager  assumed 
${\bi p}_2={\bar \alpha}{\bi F}$ with Eq.(10) and 
expressed   ${\bi E}_d$ and   $\bi F$ in terms of  
 ${\bi E}$   and ${\bi p}_1$ as     
\bea 
&&\hspace{-1cm}{\bi E}_d = \frac{\ep(\ef+2)}{2\ep+\ef}{\bi E},\\
&&\hspace{-1cm}  {\bi F}= {\bi E_d}
+ \frac{8\pi (\ep-1)(\ef+2)}{9(2\ep+\ef)}{\bi p_1},  
\ena 
See  Appendix B for more explanations of his theory. 
He  argued that the second term in Eq.(24) is parallel to 
${\bi p}_1$ and  irrelevant in orienting the dipoles 
and hence the first term ${\bi E}_d$  governs the dipole 
orientation.  Furthermore, 
let us set   ${\bi p}_i=\chi_i{\bi E}$ in equilibrium 
within Onsager's theory. 
Then, these  $\chi_1$ and $\chi_2$ 
 coincide with those in Eq.(13) (see Eq.(B5)). 
Now,  Eqs.(22) and (23) yield    
 \be
\chi_d=\chi_1 \frac{|{\bi E}|}{|{\bi E}_d|}
=  \frac{3(\epsilon-\ef)(2\epsilon+\ef)}{4\pi\epsilon (\ef+2)^2} , 
\en 
which is  the Onsager equation if 
 $\chi_d=n\mu_0^2/3k_BT$.   
From   Eqs.(13), (20), and (25) 
$\lambda_{11}$ is also calculated as\cite{Felder2}     
\be  
\lambda_{11}= \frac{4\pi}{3} - 
\frac{8\pi(\ep-1)(\ef+2)}{9(2\ep+\ef)} . 
\en  
With this expression, Felderhof's  Eq.(21) and  Onsager's 
 Eq.(24) are identical self-consistently. 
We  stress that $\chi_1$ and $\chi_2$ in Eq.(13) are 
fundamental theoretical elements, 
but use has been made of 
 different expressions for them 
in  the literature 
(see Appendix C)\cite{Fro,Zw,Hubbard,Marcus,Tomasi,Kim,Li,Lee,Mat1}. 
 
In Onsager's theory the short-range 
dipole-dipole interaction is neglected. To 
account for it, Kirkwood\cite{Kirk} set 
 \be 
\chi_d= g\mu_0^2n/3k_BT.
\en 
With $g$ in $\chi_d$, 
  Eq.(25)  is   the Kirkwood-Fr$\ddot{\rm o}$hlich (KF) 
equation\cite{Fro}. In this paper,  we define 
 $\chi_d$ by Eq.(27) with  $g$, while   
  $\lambda_{11}$ is given  by   Eq.(26)  for any  $g$. 
In Sec.IIIB,   we shall see  that 
  Onsager's results and the KF equation   
follow  from the free energy density (3).

We  make  remarks. 
(i) For $\ep\gg \ef$ we have 
$$
|{\bi E}|\sim 
|{\bi E}_d|\ll |{\bi F}| ~\sim [2\ep/3(\ef+2)]|{\bi E}|$$ 
from  Eqs.(23) and (24).    
(ii)  In equilibrium with  ${\bi p}_1=\chi_1{\bi E}$,  
 Eq.(24) is  rewritten in   the Lorentz form 
${\bi F}= [(\ep+2)/3]{\bi E}$. 
 Namely, Onsager's ${\bi F}$ in Eq.(24) 
and Lorentz's  ${\bi F}$ in Eq.(8) coincide 
for ${\bi p}_1= \chi_1{\bi E}$, but they  differ 
  for  ${\bi p}_1\neq \chi_1{\bi E}$. 
The same statement can be made for  
 Onsager's ${\bi E}_d$ in Eq.(23) 
and Felderhof's  ${\bi E}_d$ in Eq.(21). 
However, Eqs.(21) and (24) give 
the same difference  ${\bi F}-{\bi E}_d$.  
(iii) In general situations,  we should define     $\bi F$ by   Eq.(8) and 
${\bi E}_d$ by Eq.(21) or Eq.(24). 
(iv) The dipole orientation 
 is in the nonlinear regime for   
$\mu_0 |{\bi E}_d| \gs  k_BT$, which  has   been studied 
extensively\cite{Ful3,Booth,Edholm,Andel1,OnukiD}.


Let us estimate the parameters 
introduced so far. If we   set  $\ep=78.5$, $\ef=1.77$, and $\mu_0=1.85$ 
D for  ambient liquid water\cite{Edholm},  
   Eqs.(12), (13), and (25)-(27) give     
\bea 
&&\hspace{-7mm} 
\chi=6.17,~~ {\bar\alpha}=0.049, ~~
\chi_1=4.86, ~~\chi_2= 1.31,~~ \nonumber\\ 
&&\hspace{-7mm} a_{11}= 1.33, ~~\lambda_{11}= -0.95, ~~
\chi_d=2.61,~~g=2.80, \nonumber\\
&&\hspace{-7mm} n\mu_0^2/3k_BT=0.92, ~~|{\bi E}_d|= 0.070|{\bi F}|
=1.86|{\bi E}|. \nonumber
\ena

\section{Static polarization fluctuations}

In this section, we  examine  the static polarization  
correlations  treating  
${\bi p}_i =(p_{ix},p_{iy}, p_{iz})$  
as    thermal fluctuations. 
Their  wave numbers are in the intermediate 
range $\pi/H<  q<\pi/a_m$,  where $a_m$ is 
 the molecular length.  
 They  obey   the Gaussian distribution 
$\propto \exp( - {\cal F}/k_BT)$ at  $Q_0=0$.  
However,  the   variances of these 
inhomogeneous fluctuations  with $q\neq 0$ 
do not depend on  the boundary condition in 
the  linear  regime.
 
\subsection{Correlations of Fourier components}

Introducing  a  variable $\bi s$ we  rewrite  $f $  in Eq.(3)  as  
\be 
f= \frac{1}{8\pi}|{\bi E}|^2 + \frac{ 1}{2\chi}|{\bi p}|^2 + 
\frac{1}{2}A_0 |{\bi s}|^2,
 \en 
where   $\bi s$ is decoupled from ${\bi p}={\bi p}_1+{\bi p}_2$ 
and is defined by 
\be 
{\bi s}= {\sum}_{i} (a_{1i}-a_{2i}){\bi p}_i =
\frac{\ep+2}{3}\Big(\frac{{\bi p}_1}{\chi_1}-
\frac{{\bi p}_2}{\chi_2}\Big) .
\en  
We  also have     
${\bi s}= {\delta}{\cal F}/{\delta {\bi p}_1}
-{\delta}{\cal F}/{\delta {\bi p}_2}$ from Eq.(14). 
The linear response of $\bi s$ to $\bi E$ vanishes in the static limit. We    express  ${\bi p}_1$ and ${\bi p}_2$ 
in terms of $\bi p$ and $\bi s$  as 
\be 
{\bi p}_1= \frac{1}{\chi}\chi_1{\bi p} + A_0 {\bi s}, ~~~
{\bi p}_2= \frac{1}{\chi}\chi_2{\bi p} - A_0 {\bi s}.
\en 
The coefficient  $A_0$ is given  by
\be 
A_0={{\bar\alpha} \chi_1}/{\chi}
=3\chi_1\chi_2/[\chi(\ep+2)],   
\en 
which  is close to $\bar\alpha$ for $\ep\gg \ef\sim 1$.

The  Fourier components of ${\bi p}_i$ and $s$ are written as  
\be  
{\hat{\bi p}}_i({\bi q})=\hspace{-0.1mm} \int_V\hspace{-1mm}
 d{\bi r} e^{-i{\bi q}\cdot{\bi r}}
{\bi p}_i({\bi r}), ~
{\hat{ s}}({\bi q})= \hspace{-0.1mm}\int_V\hspace{-1mm}
 d{\bi r} e^{-i{\bi q}\cdot{\bi r}}
{s}({\bi r}), 
\en 
where   ${\bi q}= 2\pi(n_x/L,n_y/L,n_z/H)$ 
with $(n_x,n_y,n_z)$ being integers.  We consider  the  correlations, 
\be 
{\hat G}_{\alpha\beta}^{ij}({\bi q}) =
 \av{{\hat p}_{i\alpha} ({\bi q})
{\hat p}_{j\beta}({\bi q})^*}/Vk_BT , 
\en 
where the Greek indices refer to the Cartesian 
coordinates  $(x,y,z)$ and 
 $\av{\cdots}$ represents the equilibrium average 
at  $Q_0=0$. The inverse Fourier transformations of 
 ${\hat G}_{\alpha\beta}^{ij}({\bi q}) $ 
give the space  correlations,  
\bea 
&&\hspace{-11mm}
{ G}^{ij}_{\alpha\beta}({\bi r})
 =\hspace{-0.6mm} 
\frac{\av{p_{i\alpha}({\bi r}_1) p_{j\beta}({\bi r}_2)}}{k_BT}
=\hspace{-1.1mm}
\frac{1}{V}{\sum}_{{\bi q}}
e^{i{\bi q}\cdot{\bi r}} {\hat G}_{\alpha\beta}^{ij}({\bi q}) ,
\ena 
where ${\bi r}= {\bi r}_1- {\bi r}_2$ 
and the sum ${\sum}_{{\bi q}}$ is taken 
 in the range $\pi/H<q<\pi/a_m$.   
The positions ${\bi r}_1$ and $ {\bi r}_2$ 
are   located  in the bulk region 
(far from the boundaries), where the translational symmetry 
 nearly holds. 

We express    ${\cal F}$ 
 in  terms of   ${\hat {\bi p}}({\bi q})=
{\sum}_i {\hat {\bi p}}_i({\bi q})$ and ${\hat{ s}}({\bi q})$ as  
\be 
{\cal F}
= \frac{1}{2V}{\sum}_{{\bi q}}\Big[ 4\pi 
|{\hat{\bi p}}_\parallel({\bi q})|^2+ 
\frac{1}{\chi}|\hat{\bi p}({\bi q})|^2 
+  {A_0}|{\hat{ s}}({\bi q})|^2\Big] ,
\en 
where  the Fourier components  of ${\bi E}$ are   equated to 
   $-4\pi {\bi p}_\parallel({\bi q})$. Here, 
${\hat{\bi p}}_\parallel= 
 ({\hat{\bi q}}\cdot{\hat{\bi p}})\hat{\bi q}$ 
and ${\hat{\bi p}}_\perp={\hat{\bi p}}- {\hat{\bi p}}_\parallel$ 
are  the longitudinal and transverse  parts of ${\hat{\bi p}}$, 
respectively, with   
 $\hat{\bi q}\equiv  q^{-1}{\bi q}$ being  the unit vector 
along $\bi q$.  Then,  
\bea 
&&\hspace{-9mm} {\hat G}_{\alpha\beta}({\bi q})  
= {\sum}_{i,j}{\hat G}_{\alpha\beta}^{ij}({\bi q}) =
  \chi(\delta_{\alpha\beta}-{\hat q}_\alpha 
{\hat q}_\beta) +\frac{\chi}{\ep}  {\hat q}_\alpha 
{\hat q}_\beta , \\
&&\hspace{-9mm} \av{{\hat s}_\alpha ({\bi q}){\hat s}_\beta({\bi q})^*}
\frac{1}{V}=\frac{k_BT}{A_0} \delta_{\alpha\beta},  ~~
\av{{\hat p}_{\alpha} ({\bi q})
{\hat s}_\beta({\bi q})^*}= 0,       
\ena 
where Eq.(36)  is well known\cite{Ni,Ful1,Felder1}  
and   Eq.(37) shows that  $\bi s$ and   ${\bi p}$  are 
  {\it orthogonal} to each other.

From Eqs.(30) and (36)   we now obtain  
\bea 
&&\hspace{-12mm}
{\hat G}_{\alpha\beta}^{ij}({\bi q}) =
  (2\delta_{ij}-1) 
\delta_{\alpha\beta}  A_0 
 + (\chi_i\chi_j/\chi^2) 
{\hat G}_{\alpha\beta}({\bi q})\nonumber\\
&&  =
 \chi_{\perp}^{ij} (\delta_{\alpha\beta}- {\hat q}_\alpha{\hat q}_\beta)
+\chi_{\parallel}^{ij}{\hat q}_\alpha{\hat q}_\beta. 
\ena  
Here,  $\chi_{\perp}^{ij}$ and $\chi_{\parallel}^{ij}$ 
are   the transverse and longitudinal     variances of  
$\hat{\bi p}_i$ and $\hat{\bi p}_j$, which are  the 
inverse matrices of $a_{ij}$ and $a_{ij}'= a_{ij}+4\pi$. respectively. 
From   Eq.(7) we find  $a'_{11}a'_{22}-{a'_{12}}^2=
\ep (a_{11}a_{22}-{a_{12}}^2)$, and $a'_{22}=\ef a_{22}$. Thus,  
 \bea
&&\hspace{-8mm}
 \chi_{\perp}^{11}= 
\frac{3\chi_1}{\ef+2}= \frac{9(\ep-\ef)}{4\pi(\ef+2)^2} ,~~
\chi_{\parallel}^{11}
=\frac{\ef}{\ep}\chi_{\perp}^{11},\\
&&\hspace{-8mm} \chi_{\perp}^{12}= -\frac{\ep}{2}\chi_{\parallel}^{12}
= \frac{4\pi}{3} {\bar\alpha} \chi_1 ,\\
&&\hspace{-8mm} \chi_{\perp}^{22}- {\bar\alpha}= -\frac{\ep}{2}(
\chi_{\parallel}^{22}-{\bar\alpha})= 
\frac{4\pi}{3} {\bar\alpha} \chi_2,   
\ena 
where  we have  the sum relations, 
\be 
{\sum}_{j}\chi_{\perp}^{ij}= \chi_i,~~
{\sum}_{j}\chi_{\parallel}^{ij}= \frac{1}{\ep}\chi_i.
\en  
Here,   the longitudinal parts 
are suppressed by   the dipolar interaction. In particular,   
   $\chi_{\parallel}^{11}/\chi_{\perp}^{11}
={\ef}/{\ep}$.    From 
the trace ${\sum}_\alpha {\hat G}_{\alpha\alpha}^{11}({\bi q})$,  
we find another noteworthy relation,  
\be 
2\chi_{\perp}^{11}+ \chi_{\parallel}^{11}=
(2+\ef/\ep)\chi_{\perp}^{11}
=3\chi_d,  
\en 
where $\chi_d$ is given by Eq.(25). 
For   $\ef=1$, we simply have  $\chi_{\perp}^{11}=\chi_1=\chi$. 
For  water, we have  
 $\chi_{\perp}^{11}=3.87$ and $\chi_{\parallel}^{11}=0.087$ 
from the last paragraph in Sec.IIC.   

We  make  remarks. (i) The first line of   Eq.(38) is equivalent to  
 Felderhof's formal expression (3.11)\cite{Felder2}. 
(ii) In     Appendix C, we will present    
 ${\hat G}^{ij}_{\alpha\beta}({\bi q})$ omitting the Lorentz term. 
(iii)  Fulton\cite{Ful1} and Felderhof\cite{Felder1}  
examined  the space-time polarization correlations 
including the radiation field, where the light speed appears.

\subsection{Space correlations and $g$ factor}

We consider    the space-correlation functions  
 ${ G}^{ij}_{\alpha\beta}({\bi r})$ in Eq.(34) 
and  ${ G}_{\alpha\beta}({\bi r})= {\sum}_{i,j} 
{ G}^{ij}_{\alpha\beta}({\bi r})$.   From Eqs.(36) and (38) 
they are expressed in the range $a_m <r< H$ as 
\bea 
&&\hspace{-1cm} { G}_{\alpha\beta}^{ij}({\bi r})= 
  (2\delta_{ij}-1) 
\delta_{\alpha\beta}  A_0 \delta({\bi r}) 
 + (\chi_i\chi_j/\chi^2) 
{G}_{\alpha\beta}({\bi r}),\nonumber\\ 
&&\hspace{-1cm}
{ G}_{\alpha\beta}({\bi r}) 
= \chi\delta_{\alpha\beta}\delta({\bi r}) +  
 (4\pi\chi^2/\epsilon) \nabla_\alpha\nabla_\beta \psi(r), 
\ena 
where    $\delta({\bi r})$ represents  localized functions  
with $\int d{\bi r} \delta({\bi r})=1$, 
$\nabla_\alpha$ is  the $\alpha$ 
 component  of $\nabla$,    and $\psi(r) 
= (4\pi r)^{-1}$ for $r\gg  a_m$. 
From $\nabla^2\psi=-\delta({\bi r})$    the  traces  
(inner products)  ${\sum}_\alpha{ G}_{\alpha\alpha}^{ij}({\bi r})$ 
are short-ranged.  The   short-range behaviors   of $\delta({\bi r})$ 
and $\psi({r})$  can be known only  from microscopic simulations. 
  The expressions in Eq.(44) can be used  in the bulk and are 
independent of the boundary condition, so they are 
the correlations in  infinite systems. 
For finite systems, we should add the nonlocal terms  to 
the right hand sides of Eq.(44), which will be discussed in Sec.IV.

We suppose  a   mesoscopic sphere embedded 
in the same fluid\cite{Kirk,Bott,Fro}. Its 
  center is  at the origin $\bi 0$ in the bulk,   its 
radius $R$ is in the range $a_m\ll R\ll H$, and its 
volume $v=4\pi R^3/3$  is in the range   
$1/n\ll v\ll V$.  We integrate ${\bi p}_i({\bi r})$ 
and   ${\bi p}({\bi r})$ 
 within the sphere as     
\be
{\bi M}_i=  \int_{r<R }\hspace{-3mm} 
d{\bi r}~{{\bi p}}_i({\bi r}), ~~ 
{\bi M}={\sum}_i {\bi M}_i=  \int_{r<R }\hspace{-3mm} 
d{\bi r}~{{\bi p}}({\bi r}). 
\en  
Because  the integration region is  spherical, 
we have ${\av{ M_{i\alpha}  M_{j\beta}}}
= {\av{ {\bi M}_{i}\cdot{\bi  M}_{j}}} 
\delta_{\alpha\beta}/3$. 
Thus,  Eq.(44) gives      
\bea  
&&\hspace{-6mm}
{\av{ M_{i\alpha}  M_{j\beta}}}/{vk_BT}
 =(2\chi^{ij}_\perp + \chi^{ij}_\parallel) 
 \delta_{\alpha\beta}/3,   
 \nonumber\\
&&\hspace{-5mm}
{\av{M_\alpha M_\beta }}/{vk_BT}
={\chi} (2+1/\ep)
\delta_{\alpha\beta}/3,  
\ena 
where  we neglect  corrections of order $v/V$ 
 from the nonlocal correlations. 
We also find Eq.(46) if the integration region is a cube. 
  If it is  a  spheroid 
with the symmetry axis along   the $z$ axis\cite{Takae,Landau-e}, 
 the right-hand sides   of Eq.(46) 
become uniaxial, tending 
 to  those of Eq.(53)  in the pancake limit.  
For  $v\ll V$,   Eqs.(43) and (46) give  
\be 
{\av{|{\bi M}_1|^2}}/{3vk_BT}  
=\chi_d, 
\en
where $\chi_d$ is given in Eq.(25). 
 See   Eq.(58) for a generalization of Eq.(47) with the 
long-range correction added.

Microscopically,  ${\bi M}_1$ is 
expressed  as 
${\bi M}_1= {\sum}_k' \mu_0 {\bi \omega}_k $, 
 where we  sum  over   
the dipoles in the sphere with  
${\bi\omega}_k$ being  the unit vector 
along the  direction of the $k$-th dipole. Then, 
the $g$ factor in Eq.(27) 
 is expressed as\cite{Kirk}   
\be
g ={\av{|{\bi M}_1|^2}}/({nv\mu_0^2}) 
= 1+ {\sum}_{k\neq m}'
 \av{{\bi \omega}_k\cdot{\bi \omega}_m}/nv,  
\en  
which accounts for   the short-range orientational  
  correlation for $a_m \ll R\ll  H$. With Eqs.(47) and (48)  we recognize that 
the Felderhof free energy density (3)   can yield the KF equation.

Fr$\ddot{\rm o}$hlich\cite{Fro} expressed   
  ${\av{|{\bi M}_1|^2}}$  in his Eq.(7.38)   
 in the form of Eq.(C7), which is larger than    $\chi_d$ in  Eq.(25) 
     by a factor of  $(\ef+3)^2/9$. 
 To derive  it, he assumed ${\bi p}_2 = [(\ef-1)/4\pi]{\bi E}$ 
and ${\bi p}_1 = [(\ep-\ef)/4\pi]{\bi E}$ 
 in his Eqs.(7.35) and (7.36), which  are   invalid in the static 
limit (see  below Eq.(13)).  He then 
introduced the  effective dipole moment 
$\mu\equiv \mu_0 (\ef+2)/3$     in his Eq.(8.1) 
to   obtain  the   KF equation (25).
In contrast,   using    $\chi_1$ and $\chi_2$ in Eq.(13),   
we  do   not  modify the  dipole moment    $\mu_0$.

\subsection{Static linear response: Embedded sphere}

From Eq.(46) we  find  the 
linear response relations in terms of the variances 
among ${\bi M}_i$ in the static limit, 
\bea 
&&\hspace{-5mm}{{\bi p}_i}= \chi_i{\bi E}= 
[{\av{{\bi M}_i\cdot {\bi M}}}/{3vk_BT}] {\bi E}_c~~(i=1,2),
\nonumber\\
&& {{\bi p}}= \chi{\bi E}= 
[{\av{|{\bi M}|^2}}/{3vk_BT}] {\bi E}_c,  
\ena 
where 
$\chi_2/\chi_1 = {\av{{\bi M}_2\cdot{\bi M}}}/
\av{{\bi M}_1\cdot{\bi M}}$ follows from  Eq.(30). 
The  ${\bi E}_c$ is the  cavity field in the 
sphere\cite{Onsager,Bott,Kirk,Fro},    
\be 
{\bi E}_c= [3\ep/(2\ep+1)]{\bi E}.  
\en 
which  is produced by  the electrode charges  and the exterior 
dipoles. We can see that ${\bi E}_c$ 
is the applied field and   the interior $\bi M$ 
is its  conjugate variable.

From Eqs.(22) and (46) we write 
  ${{\bi p}_1}$  in another form,   
\be
{{\bi p}_1}=\chi_d {\bi E}_d =
[{\av{|{\bi M}_1|^2}}/{3vk_BT}] {\bi E}_d. 
\en 
where   ${\bi E}_d$ is  the   applied  field 
and its conjugate variable is 
the interior ${\bi M}_1$.  Here,   ${\bi E}_d$ 
 is  a  modified cavity field 
  including the reaction effect due to 
 the interior  ${\bi p}_2$. Thus,      
${\bi p}_2 $ cannot be 
calculated  in this scheme.    

\section{Total polarizations  and nonlocal correlations  }
 
We  consider   the total polarizations in the  cell,  
\be 
{\bi M}_i^{\rm tot}=\int_V\hspace{-1mm}
 d{\bi r}~{\bi p}_i({\bi r}),~~
{\bi M}^{\rm tot}={\sum}_i {\bi M}_i^{\rm tot}, 
\en  
which are the Fourier components, 
${\hat{\bi p}}_i({\bi 0})$ and 
${\hat{\bi p}}({\bi 0})$,  with ${\bi q}={\bi 0}$ in Eq.(32). 
The contributions from   
the polarizations in the Stern layers are negligible 
for $d\ll H$.  These homogeneous components 
 are sensitive to the boundary condition. We  introduce the nonlocal 
correlations, which are written as 
$\chi_i\chi_j D_{\alpha\beta}/\chi^2 V$ 
in $G_{\alpha\beta}^{ij}({\bi r})$ 
and $D_{\alpha\beta}/ V$ 
 in $G_{\alpha\beta}({\bi r})$ 
from Eq.(30), where the coefficients $D_{\alpha\beta}$ 
depend on the boundary condition. They   yield  
additional contributions to 
 the  variances $\av{{ M}_{i\alpha}^{\rm tot}{ M}_{j\beta}^{\rm tot}}$.  
 Some discussions on their dynamics 
will be given in   Sec.VIC. 
The case of fixed   $\Phi_a$  will be studied in  Sec.VIII.

(i)   First, setting   $Q_0=0$ with $H\ll L$,  
we integrate  
 ${ G}_{\alpha\beta}^{ij} ({\bi r}_1-{\bi r}_2)$ in    Eq.(44) 
over    ${\bi r}_1$ and ${\bi r}_2$ to find\cite{Ful3,Hansen}  
\bea 
&&
\hspace{-8mm}
{\av{M^{\rm tot}_{i\alpha}
 M^{\rm tot}_{j\beta}}}/{Vk_BT}
=\chi_{\perp}^{ij} 
(\delta_{\alpha\beta}-\delta_{\alpha z}\delta_{\beta z})
+ \chi_{\parallel}^{ij} \delta_{\alpha z}\delta_{\beta z}, 
\nonumber\\
&&\hspace{-8mm}
{\av{M^{\rm tot}_{\alpha} M^{\rm tot}_{\beta}}}/{Vk_BT}
 = \chi (\delta_{\alpha\beta}-\delta_{\alpha z}\delta_{\beta z}) 
+ \frac{\chi}{\ep}
\delta_{\alpha z}\delta_{\beta z},
\ena 
For $H\ll L$  the integration  of the dipolar term in Eq.(44) 
can be performed if use is made of the  equation,    
\be 
 \int \hspace{-1mm}d{\bi r}_\perp'   
\nabla_\alpha\nabla_\beta
  \psi(|{\bi r}-{\bi r}'|)
\cong - \delta(z-z')\delta_{\alpha z}\delta_{\beta z}, 
\en  
where  we integrate over    ${\bi r}'_\perp= (x'.y')$ and  
  $\psi(r)$  in Eq.(44)  is set equal to $1/4\pi r$.   
 Here,  Eq.(54) is obtained  from  $\int \hspace{-1mm}d{\bi r}_\perp'   
\nabla_z  \psi(|{\bi r}-{\bi r}'|) \cong - (z-z')/2|z-z'|$ 
for  $|z-z'|\ll  L$. 
Here,  the   dipolar interaction  suppresses  
the $z$ components   $M^{\rm tot}_{iz}$. 
In this case,   no nonlocal   correlation  appears 
(see the last paragraph in Sec.VIIIB).

We can derive   Eq.(53)  
even for not small $H/L$ 
if we assume    the periodic boundary condition 
along the $x$ and $y$ axes. 
To show this, we  present  another derivation of Eq.(53) 
assuming $Q_0=0$ and the lateral periodicity. In this case,  
the cell integrals of $D_z= E_z+ 
4\pi {{p}}_z$, $E_x$, and $E_y$ vanish.  
Then,   Eq.(28) yields   the free energy  from 
the Fourier components with ${\bi q}={\bi 0}$ 
in the form, 
\be 
{\cal F}_{\rm tot}
= \frac{2\pi}{V} 
|{\hat{p}}_z({\bi 0})|^2+ 
\frac{1}{2V\chi}|\hat{\bi p}({\bi 0})|^2 
+  \frac{1}{2V}A_0|{\hat{ s}}({\bi 0})|^2 .
\en 
Here,    $\hat{\bi p}({\bi 0})= {\bi M}^{\rm tot}$ and  
$\hat{\bi s}({\bi 0})= \int_Vd{\bi r}{\bi s}({\bi r})$ obey 
the distribution $\propto \exp(-{\cal F}_{\rm tot}/k_BT)$,   
so we are led  to Eq.(53).   

(ii) Second, we consider the case of   
  the periodic boundary condition along  the three   
axes\cite{Ku,Leeuw,Lada1,Lada,Bopp,Neu,Sprik}, where we have 
$\int_V d{\bi r} {\bi E}={\bi 0}$. 
Then,     the first  term in Eq.(55) is absent, leading to  
the isotropic variance relations\cite{Lebe,Neu,Leeuw,Cai}, 
\bea
&& {\av{M^{\rm tot}_{i\alpha}
 M^{\rm tot}_{j\beta}}}/{Vk_BT}= \chi_{\perp}^{ij}
\delta_{\alpha\beta},\nonumber\\
&&{\av{M^{\rm tot}_{\alpha} M^{\rm tot}_{\beta}}}
 /{Vk_BT}= \chi\delta_{\alpha\beta}~~({\rm periodic}), 
\ena 
which hold for  $L_x\times L_y\times L_z$ 
 rectangular cells.   We also find   Eq.(56) if a polar 
 fluid is enclosed  by  
an equi-potential surface or by a metal\cite{Cai,Ful4}. 
For finite $V$,  $G_{\alpha\beta}^{ij}({\bi r})$  
 consist of those in Eq.(44)  
and  the  nonlocal parts 
given by  $4\pi\chi_i\chi_j \delta_{\alpha\beta}/3\ep V$  
for   cubic cells. 
 For example, we find  
\bea 
&&{\sum}_\alpha G_{\alpha\alpha}({\bi r})=
{\av{{\bi p}({\bi r})\cdot{\bi p}({\bi 0})}}/{k_BT}  
\nonumber\\
&&= \frac{ \chi}{\ep}(2\ep+1)
\delta({\bi r})+ \frac{\chi(\ep-1)}{\ep V}   
~~({\rm periodic}),  
\ena 
where the first  term  
arises   from Eq.(44). The cell integration of Eq.(57) is $3\chi$ 
in accord with Eq.(56).  

 (iii) Third, Sprik's group\cite{Sprik,Cox} 
imposed  the global condition, 
$
\int_V d{\bi r} {\bi D}=\int_V d{\bi r} {\bi E}+
4\pi {\hat{\bi p}}({\bi 0})={\bi 0},
$ 
in their simulation. In this case,  the first term in Eq.(55) is replaced by 
${2\pi}|{\hat{\bi p}}({\bi 0})|^2/V$, so 
 the  counterpart of Eq.(56) is  obtained by replacements:  
 $\chi_{\perp}^{ij} \to \chi_{\parallel}^{ij}$ and $\chi\to \chi/\ep$.  
For cubic cells, the  nonlocal parts in 
$G_{\alpha\beta}^{ij}({\bi r})$ are 
$-8\pi\chi_i\chi_j \delta_{\alpha\beta}/3\ep V$ 
and  the second  term in Eq.(57) is   replaced by 
$-2{\chi(\ep-1)}/{\ep V}$, leading to  
 $\av{|{\bi M}^{\rm tot}|^2}/Vk_BT=3\chi/\ep$.  
 
In addition,  the nonlocal correlations 
 change the variance of the sphere integral  
${\bi M}_1$ in   Eq.(47) to\cite{Cai}   
\be 
\av{|{\bi M}_1|^2}/3vk_BT=\chi_d + C_{\rm nl}\chi_1^2v/3\ep V. 
\en  
  The Kirkwood $g$ factor should be  determined without the second term. 
The coefficient $C_{\rm nl}$ is equal to 
$ 4\pi$ for $\int_V d{\bi r} {\bi E}={\bi 0}$ 
(in the periodic case)  and to  $-8\pi$ 
for  $\int_V d{\bi r} {\bi D}={\bi 0}$, 
which  agree  with simulations\cite{Ku,Sprik}. 
In the parallel plate geometry, it vanishes  at fixed $Q_0$ 
and becomes $\ep/\chi(1+\ell_{\rm w}/H)- 1/\chi$ 
at fixed $\Phi_a$ (see Sec.VIII).

The nonlocal   correlations generally 
 appear  under global constraints.  
 In  fluid mixtures,  the space   correlations of  the 
 number densities  have nonlocal parts $\propto V^{-1}$    
in the canonical and isothermal-isobaric 
 ensembles\cite{Lebo, OnukiP}, which  
 do not exist   in the grand-canonical ensemble.
For example,  the pair   correlation of the density 
fluctuation $\delta {\hat n}({\bi r})$ in pure fluids behaves 
in the canonical ensemble  as 
\be 
\av{\delta {\hat n}({\bi r})\delta {\hat n}({\bi 0})}= 
{ n}\delta({\bi r})+{n}^2 [g(r)-1]-
 {n}^2k_BT\kappa_T/V,
\en  
 Here,  $n$ is the  mean density, 
  $g(r)$ is the radial 
distribution function,  and  
$\kappa_T$ is the  isothermal  compressibility. 
The cell  integration of Eq.(59)  
vanishes due to the thermodynamic relation 
 $ {n}k_BT\kappa_T=1+ { n}\int d{\bi r}[g(r)-1]$.

\section{Solvation free energy with a charge density}

Marcus\cite{Marcus} studied the  electron 
transfer kinetics  in  a polar solvent,  where 
the solvent polarization around ions was assumed to obey the 
classical electrostatics.   In this approach, various  
     chemical reactions in a polar solvent have been 
studied\cite{Tomasi,Kim,Li,Mat1,Fleming,Bagchi,N1,M2}. 
Marcus  wrote  the fastest 
 electronic polarization as ${\bi p}_e$ 
and the  sum of the slower  orientational 
and atomic   ones as ${\bi p}_u$. He then  expressed    ${\bi p}_e$  
in terms of   ${\bi p}_u$ 
and the {\it bare  electric field} ${\bi E}_0$ produced by 
the solute charge density $\rho_{\rm s}$ and the electrode charges 
in local equilibrium.  
In these     papers, however,    
the Lorentz term in the free energy 
has been missing. Hence,   it is included 
in this section.  
In our theory,   ${\bi p}_1$ denotes the orientational polarization 
and  ${\bi p}_2$  the sum of the atomic and electronic ones. 
The  relations here 
 will be used in Secs.VI-VIII.

The bare field  ${\bi E}_0$ 
does not include   the polarization contribution and   
 is  determined   by\cite{Marcus,Lee}    
\be 
{\bi E}_0= -\nabla\Phi_{\rm bare}, ~~~\nabla\cdot{\bi E}_0= 
-\nabla^2\Phi_{\rm bare} =4\pi{\rho}_{\rm s}, 
\en  
where   $\Phi_{\rm bare}$ is  the bare 
potential and   the electrode  charges 
appear in the boundary conditions. 

We   divide ${\bi p}_i({\bi r})$ into 
the longitudinal part ${\bi p}_{i{\parallel}}({\bi r})$ 
and the transverse part ${\bi p}_{i{\perp}}({\bi r})
$, where their  Fourier components are 
${\hat{\bi p}}_{i{\parallel}}({\bi q}) $ 
and  ${\hat{\bi p}}_{i{\perp}}({\bi q}) $, respectively 
(see below Eq.(35)). 
If ${\bi p}_i$ depends only on $z$, 
we have ${\bi p}_{i\parallel}=(0,0, { p}_{iz})$.
From      $\nabla\cdot{\bi D}=4\pi\rho_{\rm sol}$ we find   
\be 
{\bi E}_0= {\bi E}+ 4\pi {\bi p}_{\parallel}=
 {\bi D}-4\pi{\bi p}_{\perp},
\en   
where  ${\bi p}_{\parallel}= {\bi p}-{\bi p}_{\perp}= 
{\bi p}_{1\parallel}+{\bi p}_{2\parallel}$. The fields  ${\bi E}_0$ and $ {\bi E}$  are 
longitudinal. 
Thus, ${\bi E}_0=  {\bi D}$ for ${\bi p}_{\perp}={\bi 0}$. 

Minimization of   ${\cal F}$   with respect 
to ${\bi p}_2$ at fixed ${\bi p}_1$ and   $Q_0$ is attained 
at     ${\bi p}_2={\bi  p}_2^{\rm eq}$ as in  Eqs.(6) and (9). 
Using Eq.(61) we  rewrite   this ${\bi  p}_2^{\rm eq}$ 
 in terms of  ${\bi p}_1$ 
and   ${\bi E}_0$ as  
\be 
{\bi p}_{2}^{\rm eq} 
 = \frac{\ef-1}{4\pi\ef} \Big[ {\bi E}_0
-\frac{8\pi}{3} {\bi p}_{1\parallel}\Big] 
+\frac{\ef-1}{3} {\bi p}_{1\perp}.
\en 
Then,   ${\cal F}$  
at  ${\bi p}_2={\bi p}_2^{\rm eq}$ 
is written as 
\bea 
&&\hspace{-10mm} {\cal F}_{\rm s}=
\int_V\hspace{-1mm} d{\bi r}\Big[\frac{|{\bi E}_0|^2}{8\pi\ep}+ 
\frac{1}{2\chi_{\parallel}^{11}}
 \Big |{\bi p}_{1{\parallel}}- {{\bi p}}_1^{\rm eq}\Big |^2+  
\frac{|{\bi p}_{1\perp}|^2}{2\chi_{\perp}^{11}}
  \Big], 
\ena 
where   $\chi_{\perp}^{11}$ and $\chi_{\parallel}^{11}$ 
are given  in  Eqs.(39) and (43) and  
\be 
{\bi p}_1^{\rm eq}= \frac{\chi_1}{\ep}{\bi E}_0.
= \frac{3(\ep-\ef)}{4\pi(\ef+2)\ep} {\bi E}_0. 
\en  
Here,   ${\bi p}_{1{\parallel}}$ and ${\bi E}_0$ 
evolve slowly, to which 
${\bi p}_{1\perp}$ is uncoupled  in 
the linear order.

We are interested in the thermal  fluctuation of 
  ${\bi p}_{2}$    around 
${\bi p}_2^{\rm eq}$ in Eq.(62), so  we consider its     deviation,   
\be 
{\bi \xi}=  {\bi p}_2-{\bi p}_2^{\rm eq},  
\en   
which will be important in the next section. 
Retaining $\bi \xi$ 
we express  ${\bi E}$ in Eq.(61) and ${\bi p}= {\bi p}_1+ 
{\bi p}_2$  as 
\bea 
&&\hspace{-1cm} {\bi E}=\frac{1}{\ef} \Big[{\bi E}_0 
 -\frac{4\pi}{3}  (\ef+2) {\bi p}_{1\parallel}\Big] -
4\pi{\bi \xi}_\parallel,\\
&&  \hspace{-1cm}
{\bi p}=  \frac{\ef-1}{4\pi\ef} {\bi E}_0+ 
\frac{\ef+2}{3\ef} 
\Big( {\bi p}_{1\parallel} + 
\ef{\bi p}_{1\perp}\Big)+{\bi \xi}.
\ena 
 The free energy 
increase due to  $\bi \xi$ is calculated as 
\be 
{\cal F}_\xi={\cal F} -{\cal F}_{\rm s}
=\frac{1}{2}{a_{22}}\int_V d{\bi r}
\Big[\ef |{\bi \xi}_{{\parallel}} |^2
+ |{\bi \xi}_{{\perp}} |^2   \Big],  
\en 
where  ${\bi \xi}_\parallel$ and ${\bi \xi}_\perp$ are  
 the longitudinal and transverse parts  of ${\bi \xi}$, respectively. 
Thus, 
 $\int d{\bi r}\av{\xi_\alpha({\bi r})p_{1\beta}({\bi 0})}=0$, 
so $\bi \xi$ is {\it orthogonal} to  ${\bi p}_1$. 
 From the above  ${\cal F}_\xi$ we obtain 
the variances of    the Fourier components 
 ${\hat{\bi \xi}} ({\bi q}) $ 
 of ${\bi \xi}({\bi r})$:    
\bea 
&&\hspace{-10mm}
\frac{\av{{\hat\xi}_{\alpha} ({\bi q})
{\hat\xi}_{\beta}({\bi q})^*}}{Vk_BT}
 = \frac{\ef-1}{4\pi}\Big[(\delta_{\alpha\beta}-
{\hat{q}}_\alpha {\hat{q}}_\beta)  
+  \frac{{\hat{q}}_\alpha {\hat{q}}_\beta}{\ef} \Big]. 
\ena 
In $\bi E$ in Eq.(66), the last term  
$-4\pi {\bi \xi}_\parallel$  
 serves  as a rapidly varying random noise 
(see Eqs.(74) and (75)).

Lee and Hynes\cite{Lee} presented the solvation 
free energy   ${\cal F}_{\rm s}$ 
in their Eq.(2.4), which is 
equivalent to that of Marcus\cite{Marcus}. 
Let us  rewrite the second term in  Eq.(63) as 
\be 
\frac{1}{2\chi_{\parallel}^{11}}
 \Big |{\bi p}_{1{\parallel}}- {{\bi p}}_1^{\rm eq}\Big |^2=
\frac{\ep-\ef}{8\pi\ep\ef}\Big| {\bi E}_0- \frac{\ef+2}{3}\cdot 
\frac{4\pi\ep{\bi p}_{1\parallel}}{\ep-\ef}\Big|^2, 
\nonumber
\en 
where use is made of Eq.(39).  
We can see  that the factor $(\ef+2)/3$ in the above 
expression does not appear in  Lee-Hynes'   ${\cal F}_{\rm s}$. 
In  Appendix C, we will give more 
 comments on the previous linear theories.

\section{Polarization dynamics }
\subsection{Dynamic equations}  

We now investigate   the linear 
dynamics of the time-dependent 
polarizations ${\bi p}_i ({\bi r},t)$ 
at  long wavelengths, where the translational  motions are  negligible. 
The temperature $T$ and the dipole   density $n$ are 
 homogeneous constants. 
We set up  simple  relaxation equations\cite{Landau-s,Onukibook}, 
\bea 
&&\hspace{-10mm}
\frac{\p }{\p t} {\bi p}_1= -L_1 \frac{\delta }{\delta{\bi p}_1}
{\cal F}
=  \frac{\chi_1}{\tau_{\rm L}\ep}{\bi E}_0-
\frac{1}{\tau_{\rm L}} {\bi p}_{1{\parallel}}- 
  \frac{1}{\tau_{\rm D}}{\bi p}_{1\perp} 
\nonumber\\
&&\hspace{2mm}- ({4\pi}/{3})L_1 (2{\bi \xi}_\parallel- {\bi \xi}_\perp), 
\\
&&\hspace{-10mm}
\frac{\p }{\p t} {\bi p}_2= -L_2 \frac{\delta }{\delta{\bi p}_2}
{\cal F}= - \frac{1}{\tau_{\rm f}}(\ef{\bi \xi}_{\parallel}
+{\bi \xi}_\perp) ,
\ena   
where   $L_1$ and $L_2$ are kinetic coefficients
  with   $L_1\ll  L_2$  and  
  $\delta {\cal F}/\delta {\bi p}_i= {\sum}_j 
a_{ij} {\bi p}_j-{\bi E}$ at fixed $Q_0$. 
We should replace   $ {\cal F}$  by 
 $\tilde{\cal F}$ in Eq.(15) at fixed $E_a$ to 
obtain the same equations (see  below Eq.(16)). 
The right hand sides are expressed  in terms of 
  ${\bi p}_1$, $\bi \xi$, and ${\bi E}_0$ using  
${\cal F}_{\rm s}$   and $ {\cal F}_\xi$ in  Eqs.(63) and (68).
For  stationary ${\bi E}_0$,
we have 
$$
\frac{d}{dt} {\cal F}= - {\sum}_i L_i 
|\delta {\cal F}/\delta {\bi p}_i|^2\le 0,
$$  so     the equilibrium 
determined by $\delta {\cal F}/\delta {\bi p}_i={\bi 0}$ 
is approached as $t\to \infty$. 
   

In Eqs.(70) and (71) 
we define  $\tau_{\rm D}$, $\tau_{\rm L}$, and $\tau_{\rm f}$  as   
\bea 
&&\hspace{-10mm} \tau_{\rm D}= \chi^{11}_{\perp}/L_1, ~~~ 
\tau_{\rm L}= \chi^{11}_{\parallel}/L_1= \tau_{\rm D} \ef/\ep,\nonumber\\
&&\tau_{\rm f}= (\ef-1)/(4\pi  L_2). 
\ena    
At fixed ${\bi E}_0$,   $\tau_{\rm D}$ 
is  the relaxation  time of ${\bi p}_{1\perp}$ and  
 $\tau_{\rm L}$ is that of  ${\bi p}_{1\parallel}$, 
while   $\tau_{\rm f}$ is that of ${\bi \xi}$. The $\tau_{\rm D}$  is the 
Debye relaxation time in Eqs.(1) and (89) below. 
The dynamical relation $ \tau_{\rm L}/\tau_{\rm D}=\ef/\ep$  
is well known in the 
literature\cite{Elton,Neu,
Hubbard,Berne,Fleming,Mad,Bagchi,Deutch,Mukamel}.
In our theory, it follows from   
 the static relation 
   $\chi^{11}_{\parallel}/\chi^{11}_\perp=
\ef/\ep$ in Eq.(39). 
In the limit     $\tau_{\rm f} /\tau_{\rm L}\to 0$, we 
are led to ${\bi \xi}={\bi p}_2- {\bi p}_2^{\rm eq}={\bi 0}$. 
The relaxation equation (67) is  
convenient to take this limit in the time-correlation 
functions below, though its Markovian form 
 is a  very  crude approximation for microscopic 
$\tau_{\rm f}$.

Using  Eqs.(13), (64), and (66), 
we can rewrite Eq.(70) into    Hubbard-Onsager's Eq.(2.9)\cite{Hubbard}: 
\be 
{\tau_{\rm D}}\frac{\p {\bi p}_1}{\p t}
 = \chi_1{\bi E}-{\bi p}_1 ,
\en  
where we set  ${\bi \xi}={\bi 0}$ in Eq.(70).   
However, these authors assumed   $\chi_1=  
(\ep-\ef)/4\pi$ and $\chi_2= 
(\ef-1)/4\pi$, which also lead  to  
$ \tau_{\rm L}/\tau_{\rm D}=\ef/\ep$ (see  Appendix C).

As  is well  known,   a  moving ion and  
  a  rotating dipole are exerted  by  a  
relaxing  electric field produced  by the surrounding 
dipoles,  resulting in  a  dielectric 
friction\cite{Mason,Mad,Wo,Boyd,Zw,Zw1,Hubbard,Felder3,Zwan,Mat2}, 
If this effect is included,  $L_1$ in Eq.(70) can be 
frequency-dependent on the scale of $\tau_{\rm L}^{-1}$. 
In this paper,   we treat $L_1$ as a constant  as a first step.   

  We should also generalize Eqs.(70) and (71) 
 accounting for the diffusion 
and the convection at finite $q$\cite{Hubbard,Zw}. 
 Some authors\cite{Ni,Wolynes,Bagchi,Wo} 
 examined  the dynamics of   
the position-angle distribution 
 $\rho({\bi r},{\bi \omega},t)$ of the dipoles  
for $\ef=1$.

\subsection{Short-range time-correlation functions}

We calculate the polarization time-correlations 
at  long wavelengths with  ${\bi E}_0={\bi 0}$ in the bulk 
region,  neglecting  the nonlocal correlations. 
In this case,  we should treat  Eqs.(70) and (71) as Langevin 
equations\cite{Onukibook,Landau-s} 
describing the polarization dynamics, though we do not write 
 the    random source terms for simplicity.

First, we consider   the time-correlation  
of ${\bi \xi}$ in Eq.(65):  
\be 
G_{\rm f}(t)= \hspace{-1mm}\int\hspace{-1mm} 
d{\bi r}\frac{
{\av{{{\bi \xi}} ({\bi r},t)\cdot 
{{\bi \xi}}({\bi 0},0)}}}{3k_BT}=
\frac{2}{3}G^{\rm f}_{\perp}(t)
+ \frac{1}{3}G^{\rm f}_{\parallel}(t),
\en 
where $G^{\rm f}_{\perp}(t)$ arises from 
${\bi \xi}_\perp$ and $ G^{\rm f}_{\parallel}(t)$ 
from ${\bi \xi}_\parallel$. 
For $0<t\ls \tau_{\rm f}$,    ${\bi p}_1$ is 
unchanged and     $\p {\bi p}_2/\p t$ 
can be  equated  to  $ \p {\bi \xi}/\p t$ in  Eq.(71). 
Thus,  Eq.(71) is integrated to give  
\bea
&&\hspace{-10mm}
 G^{\rm f}_{\perp}(t)= \frac{\ef-1}{4\pi}e^{-{t}/{\tau_{\rm f}} },~~ 
G^{\rm f}_{\parallel}(t)= \frac{\ef-1}{4\pi\ef} e^{-{\ef t}/{\tau_{\rm f}}} ,
\ena  
which decay   rapidly with  $\tau_{\rm f}$. However, 
 in  experiments and 
simulations\cite{Fleming,Roy,Lada1,Lada},  
 the polarization time-correlations 
decayed   non-exponentially    
at short times.

\begin{figure}[t]
\includegraphics[scale=0.6]{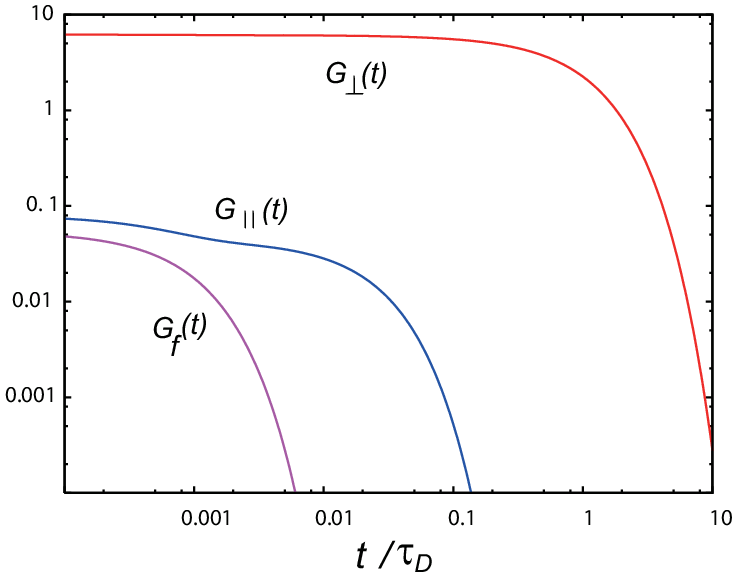}
\caption{\protect  
 $G_{f}(t)$, 
$G_{\perp}(t)$, and $G_{\parallel}(t)$ vs $t/\tau_{\rm D}$ 
on a log-log scale for  water, where  
$\ep=78.5$, $\ef=1.77$, and $\tau_{\rm f}/\tau_{\rm D}=10^{-2}$. 
In this case  $G_\perp(t)$ is much larger 
than the others and $G(t)\cong 2G_\perp(t)/3$. 
}
\end{figure}

Second,  Eq.(70) gives   the  
time-correlation of ${\bi p}_1$: 
\bea 
&&\hspace{-1cm}{ G}_{\rm or}(t) =
\int d{\bi r}  \av{{{\bi p}}_1 ({\bi r},t)\cdot
{{\bi p}}_{1}({\bi 0},0)} /3k_BT\nonumber\\
&& =2\chi_\perp^{11}  e^{-t/\tau_{\rm D}}+
\chi_\parallel^{11}  e^{-t/\tau_{\rm L}},  
\ena 
where  $G_{\rm or}(0)= \chi_d$ from Eq.(43). 
This function is continuous at $t=0$ even 
in the limit  $\tau_{\rm f}\to 0$.

Third, we examine the time-correlation  of ${\bi p}$:  
\bea 
&&\hspace{-10mm}
{ G}(t) =\int d{\bi r}  \av{{{\bi p}} ({\bi r},t)\cdot
{{\bi p}}({\bi 0},0)}/{3k_BT} \nonumber\\
&&\hspace{-3mm}  =   G_{\rm f}(t) + \frac{\ep-\ef}{12\pi} \Big[ 
2 e^{-t/\tau_{\rm D}}+
\frac{1}{\ep\ef} e^{-t/\tau_{\rm L}}\Big] ,
\ena 
where  $G(0)= \chi(2+1/\ep)/3$ and use is made of   
Eq.(67).   Since $G_{\rm f}(t)$ decays rapidly, 
  $G(t)$ decreases in the time interval $0<t\ls \tau_{\rm f}$     by 
$G_{\rm f}(0)= (\ef-1)(2+1/\ef)/{12\pi}. 
$

In ${ G}_{\rm or}(t)$ and ${ G}(t)$, 
the transverse parts are   much 
larger than the longitudinal parts   for $\ep\gg \ef$.  
We write the  transverse and   longitudinal parts of 
 $G(t)$   as   $G_{\perp}(t)$ and   
$G_{\parallel}(t)$, respectively. Some calculations give  
\bea
&&\hspace{-1cm}G_{\perp}(t) =
\int d{\bi r}  \av{{{\bi p}_\perp} ({\bi r},t)\cdot
{{\bi p}}_\perp({\bi 0},0)}/2k_BT,\nonumber\\
&&=  G^{\rm f}_\perp(t)+  [(\ep-\ef)/4\pi]   e^{-t/\tau_{\rm D}},\\
&&\hspace{-1cm}G_{\parallel}(t) = 
\int d{\bi r}  \av{{{\bi p}_\parallel} ({\bi r},t)\cdot
{{\bi p}}_\parallel({\bi 0},0)}/k_BT\nonumber\\
&& = G^{\rm f}_\parallel(t)+ [(\ep-\ef)/4\pi\ep\ef]  e^{-t/\tau_{\rm L}}.
\ena
where  $G(t)= [2G_{\perp}(t)+G_{\parallel}(t)]/3$,  
$G_{\perp}(0)=\chi$, and 
 $G_{\parallel}(0)=\chi/\ep$. 
 Thus, $G_{\perp}(t)$ and $G_{\parallel}(t)$ decrease  by 
$G_\perp^{\rm f}(0)$ and $G_\parallel^{\rm f}(0)$ 
in the time range $t\ls \tau_{\rm f}$ and subsequently  decay with 
$\tau_{\rm D}$ and $\tau_{\rm L}$, respectively. 
 Previously, Madden and  Kivelson\cite{Mad} predicted 
the long-time behaviors of $G_\perp(t)$ and $G_\parallel(t)$ 
for $\ef=1$. In Fig.2, we plot $G_{f}(t)$, 
$G_{\perp}(t)$, and $G_{\parallel}(t)$ vs $t/\tau_{\rm D}$ 
for  water. 

Furthermore, at  finite wave number $q$, 
we can introduce the $q$-dependent time-correlation 
functions,
\bea 
&&G_{\perp}(q,t)=  
{\av{{\hat{\bi p}}_\perp({\bi q},t)
\cdot{\hat{\bi p}}_\perp({\bi q},0)^*}}/{2Vk_BT},\nonumber\\
&&G_{\parallel}(q,t)={ \av{{\hat{\bi p}}_\parallel({\bi q},t)
\cdot{\hat{\bi p}}_\parallel({\bi q},0)^*}}/{Vk_BT},
\ena 
where  ${\hat{\bi p}}_\perp({\bi q},t)$ and 
${\hat{\bi p}}_\parallel({\bi q},t)$ are 
the transverse and longitudinal parts of 
the Fourier component ${\hat{\bi p}}({\bi q},t)$, respectively. 
To find the $q$-dependence of these functions, however,  
we should perform microscopic simulations 
or introduce  the gradient free 
energy\cite{Onukibook,Maggs,Ko,Ma}. 
Then, the  time-correlation of  ${\hat{\bi p}}({\bi q},t)$ 
is  written as  
\bea 
&&\hspace{-15mm}{\hat G}_{\alpha\beta}({\bi q},t) 
=\av{{\hat p}_\alpha({\bi q},t){\hat p}_\beta({\bi q},0)^* }/Vk_BT 
\nonumber\\
&&\hspace{-1cm}=
 G_{\perp}(q,t) (\delta_{\alpha\beta} - {\hat q}_\alpha{\hat q}_\beta)+  
G_{\parallel}(q, t) {\hat q}_\alpha{\hat q}_\beta. 
\ena 
 In the small $q$ range 
$q\ll a_m^{-1}$ we have  $G_{\perp}(q,t) \cong G_{\perp}(t)$  and 
$G_{\parallel}(q, t)\cong G_{\parallel}(t)$. 
Thus, Eq.(81) yields  
the space-time correlation for  $ a_m\ll r< H$:   
\bea 
&&\hspace{-10mm}{ G}_{\alpha\beta}({\bi r},t) 
=\av{p_\alpha({\bi r},t) p_\beta({\bi 0},0) }/k_BT 
\nonumber\\
&&\hspace{-1cm}=
 G_{\perp}(t) \delta_{\alpha\beta}\delta({\bi r}) +  
[ G_{\perp}(t) -G_{\parallel}(t)] 
  \nabla_\alpha\nabla_\beta {\psi(r)}, 
\ena
which tends to $G_{\alpha\beta}({\bi r})$ in Eq.(44) as $t\to 0$ 
and gives  ${\sum}_\alpha{ G}_{\alpha\alpha}({\bi r},t) 
= 3G(t)\delta({\bi r})$  
in accord with Eq.(77).

In the above time-correlation functions,  
we take  the inner products of the two vectors, 
for which the long-range 
dipolar correlation is cancelled (see   below Eq.(44)).  
We  integrate them in a region 
 much smaller than $V$ to neglect the nonlocal 
correlations.  If the  integration   is  
in  a sphere  $r<R$, the integrals are well defined 
for  $R_c<R\ll H$, where  $R_c\sim  1$ nm 
in  massive simulations  by Alvarez {\it et al.}\cite{Alva}  
(with  dipole  numbers  $\gs 10^6$). 

\subsection{Time correlations of total polarizations}
In Sec.IV we have calculated   the variances of the  
 total polarizations ${\bi M}_i^{\rm tot}$.
Here, we   consider their   time-correlation functions. 
(i) At  fixed $Q_0$, we integrate  
  Eq.(82) in the cell using Eq.(54) to find  
\bea 
&&\hspace{-8mm} 
\av{M_\alpha^{\rm tot}(t)
M_\beta^{\rm tot}(0)}/Vk_BT=
\int_V d{\bi r}G_{\alpha\beta}({\bi r},t) \nonumber\\
&& =
 G_\perp(t)(\delta_{\alpha\beta} -\delta_{\alpha z}
\delta_{\beta z})
+ G_\parallel(t)\delta_{\alpha z}
\delta_{\beta z} , 
\ena   
which tends to the last equation in Eq.(53) as $t \to 0$. 

(ii) Second,  in the periodic boundary condition,
 we have     $\int_V d{\bi r}{\bi E}({\bi r},t)
={\bi 0}$, so  Eqs.(73) and (78) give 
\bea 
&&\hspace{-10mm} \av{{ M}_{1\alpha}^{\rm tot}(t) {M}_{1\beta}^{\rm tot}(0)}
/Vk_BT=     \delta_{\alpha \beta} \chi_{\perp}^{11}
 e^{-t/\tau_D},\nonumber\\
&&\hspace{-10mm}
\av{{ M}_{\alpha}^{\rm tot}(t) {M}_{\beta}^{\rm tot}(0)}
/Vk_BT =  \delta_{\alpha \beta}G_\perp(t) ~({\rm periodic}).
\ena  
which tend to those in Eq.(56) as $t\to 0$. We will 
relate  $G_\perp(t)$ to $\ep^*(\omega)$ in Eq.(97). 
In some simulations in the periodic boundary 
condition\cite{Lada1,Alva,Ku},  
$\av{{\bi M}^{\rm tot}(t)\cdot {\bi M}^{\rm tot}(0)}$ 
 decayed with $\tau_D$, which 
slightly deviated from the single-exponential Debye form 
due to dipole librations. 

(iii) Third, in the Sprik 
condition $\int_V d{\bi r}{\bi D}({\bi r},t)=
{\bi 0}$,     
 the counterpart  of Eq.(84) is obtained  by replacements:  
$\chi_{\perp}^{11} e^{-t/\tau_D}\to  \chi_{\parallel}^{11} e^{-t/\tau_L}$  
and $G_\perp(t)\to G_\parallel(t)$, 
which are consistent  
 with the simulation\cite{Sprik1}.

\section{Dynamical linear response }

\subsection{Frequency-dependent dielectric functions} 

In this subsection, 
 we  examine the dielectric response 
to an  applied field oscillating with frequency $\omega$ 
in the  geometry of parallel metal plates. 
The    deviations  are   the statistical  averages 
and   are written in complex numbers 
depending  on time as  $e^{i\omega t}$,  
where their real parts have their physical meaning.

The  vector deviations  are    written as 
 \be 
{\bi p}_i = p_i^* {\bi e}_z, ~~~
{\bi p} = p^* {\bi e}_z, ~~~
{\bi E}_0 = E_0^*  {\bi e}_z,~~ 
{\bi E} = E^* {\bi e}_z, 
\en 
where ${\bi e}_z$ 
is the unit vector along the $z$ axis. 
Here,  $p_i^*$ and $E^*$  are homogeneous  in the bulk 
 but   depend on $z$ near  the walls. 
We relate    ${ E}_0^*$  and  ${ E}^*$  as 
\be 
 E_0^* = \ep^*(\omega) E^* =  4\pi {\bar \sigma}_0\propto e^{i\omega t}. 
\en 
where   $ \ep^*(\omega)=1+4\pi  \chi^*(\omega)$ is 
   the  frequency-dependent dielectric constant. 
We define  the    frequency-dependent  
dielectric susceptibilities for ${\bi E}$ and ${\bi E}_0$ as 
\be 
\chi_i^*(\omega)=p_i^*/E^*, ~~\alpha_i^*(\omega)= p_i^* /E_0^*.
\en  
Their sums are   the net susceptibilities  
$ \chi^*(\omega) =
p^*/E={\sum}_i \chi_i^*(\omega)$ and  
$\alpha^*(\omega)= p^*/E_0={\sum}_i \alpha_i^*(\omega)$, where 
\be 
\alpha_i^*(\omega)={\chi_i^*(\omega)}/{\ep^*(\omega)}, 
~~ \alpha^*(\omega)=[1-1/\ep^*(\omega)]/4\pi .
\en  

In the present  situation, ${\bi p}_i$
 are longitudinal (or ${\bi p}_{i\perp}={\bi 0}$). 
Setting   ${\bi \xi}={\bi 0}$, 
 we have  $(1+i\omega\tau_{\rm L})\alpha_1^*=
\chi_1/\ep$ from Eq.(70) and $\alpha_2^*=
(1-1/\ef)(1- 8\pi \alpha_1^*/3)/4\pi$
from  Eq.(62). Then, we obtain   
all the  susceptibilities  as 
\bea 
&&\hspace{-12mm} \ep^*(\omega) 
=\ep \frac{1+i\omega\tau_{\rm L}}{1+i\omega\tau_{\rm D}}
= \ef + \frac{\ep-\ef}{1+ i\omega\tau_{\rm D}}, \\ 
&&\hspace{-12mm}\chi_1^*(\omega)
= \frac{\chi_1}{1+i\omega \tau_{\rm D}}
= \frac{3}{4\pi} \cdot\frac{\ep^*(\omega)-\ef}{\ef+2},
\\
&&\hspace{-12mm} \chi_2^*(\omega)=
 \frac{\ef-1}{4\pi}
\Big[1+ \frac{4\pi}{3} \chi_1^*(\omega)\Big], \\
&&\hspace{-12mm} \alpha^*(\omega)
= \frac{\ef-1}{4\pi\ef}+ \frac{\ep-\ef}{4\pi\ep\ef
(1+i\omega\tau_{\rm L})},\\
&&\hspace{-12mm} \alpha_1^*(\omega)= 
 \frac{\chi_1}{\ep(1+ i\omega\tau_{\rm L})}
= \frac{3}{4\pi} \cdot\frac{\ep-\ef}{(\ef+2){\ep(1+ i\omega\tau_{\rm L})}},
\\
&&\hspace{-12mm}
\alpha_2^*(\omega)=
\frac{\ef-1}{4\pi\ep\ef}\Big[\ep- 
\frac{8\pi\chi_1}{3(1+ i\omega\tau_{\rm L})}\Big], 
\ena 
which  hold  for $\omega\tau_{\rm f}\ll 1$. Note that  
     Eq.(89)  is the Debye formula (1). 
Here,   $\ep^*(\omega)$ and $\chi_i^*(\omega)$ 
are characterized by      $\tau_{\rm D}$,  while 
 $\alpha^*(\omega)$ and $\alpha_i^*(\omega)$  by     $\tau_{\rm L}$.
See  their relaxation functions 
in  Eqs.(97)-(102) below. 

For $\ep\gg \ef$ and $\ef-1\gs 1$  we notice   that   
$\ef$ has the meaning of  the high-frequency 
dielectric constant  in the frequency
 range $\tau_{\rm L}^{-1}\ll \omega\ll 
\tau_{\rm f}^{-1} $, where $|{\bi p}_2|\gg |{\bi p}_1|$ and   
\be 
\chi^*(\omega)  \cong  \chi_2^*(\omega)\cong (\ef-1)/4\pi . 
\en     
See the sentences below Eq.(13) on the static limit $\chi_2$. 

 In addition, we note that Eq.(70) and (71) 
can  be solved  for general  $\omega$ under Eq.(85). 
 For example, in the case   
$L_1/L_2\ll 1$, $\ep^*(\omega)$ and $\alpha^*(\omega)$ 
are calculated as   
\bea 
&& \hspace{-10mm} 
\ep^*(\omega)-1=   \frac{\ef-1}{1+ i\omega\tau_{\rm f}}
 + \frac{\ep-\ef}{1+ i\omega\tau_{\rm D}},\nonumber\\
&& \hspace{-10mm} 
4\pi\alpha^*(\omega)=   \frac{\ef-1}{\ef+ i\omega\tau_{\rm f}}
 + \frac{\ep-\ef}{\ep\ef(1+ i\omega\tau_{\rm L})},  
\ena 
where the first terms involve $\tau_{\rm f}$.

\subsection{Relaxation functions and previous theories} 

We express   $\chi^*(\omega) $ and $\alpha^*(\omega) $ 
in Eqs.(89) and (92)  in terms of  $G_{\perp}(t)$  and $G_{\parallel}(t)$ 
in Eqs.(78) and (79) as\cite{Kubo}      
\bea 
&&\hspace{-10mm} 
\chi^*(\omega) =p^*/E^*=
G_{\perp}(0) - i\omega\int_0^\infty
 dt e^{-i\omega t} G_{\perp}(t),\\
&&\hspace{-10mm} \alpha^*(\omega)=p^*/E^*_0 
=G_{\parallel}(0)  -i\omega\int_0^\infty dt e^{-i\omega t}G_{\parallel}(t). 
\ena 
Here, the right hand sides  are   the Fourier-Laplace 
(FL)  transforms  of  the {\it response  functions}
 $- dG_{\perp}(t)/dt$ and  
$- dG_{\parallel}(t)/dt$, while 
   $G_{\perp}(t)$ and $G_{\parallel}(t)$ 
are  the {\it relaxation functions}\cite{Kubo}. 
We derive  these  relations using  the explicit 
expressions of $\chi^*$, $\alpha^*$, $G_\perp$, 
and $G_\parallel$.  Previously,  they were derived    
in the Hamiltonian formalism\cite{Mukamel,Ful2,Deutch,Mad,Bagchi}. 
See   Appendix D  for more discussions on our scheme.

We  can also express 
 $\chi_i^*(\omega)$  and $\alpha_i^*(\omega)$ 
in the same manner  by replacements:   
 $G_{\perp}(t)\to G_{i\perp}(t)$ in Eq.(97) 
 and   $G_{\parallel}(t)\to  
G_{i\parallel}(t)$ in Eq.(98), respectively, 
where we define 
\bea 
&&\hspace{-10mm} G_{i\perp}(t)\equiv \int d{\bi r}  
\av{{\bi p}_{i\perp} ({\bi r},t)\cdot
{\bi p}_\perp({\bi 0},0)}/2k_BT,  \\
&&\hspace{-10mm} G_{i\parallel}(t) = 
 \int \hspace{-1mm} d{\bi r}
\av{{\bi p}_{i\parallel}({\bi r},t)\cdot 
{{\bi p}_\parallel ({\bi 0},0)}}/k_BT.
\ena 
Here, ${\sum}_i G_{i\perp}(t)=G_{\perp}(t)$,  
${\sum}_i G_{i\parallel}(t)=G_{\parallel}(t)$, 
$G_{i\perp}(0)= \chi_i$, and  $G_{i\parallel}(0)= \chi_i/\ep$. 
For $t\gg \tau_{\rm f}$ we find   
\bea 
&&\hspace{-12mm}
G_{1\perp}(t) = \chi_1 e^{-t/\tau_{\rm D}},~~
G_{2\perp}(t) =\frac{\ef-1}{3} \chi_1 e^{-t/\tau_{\rm D}},\\
&&\hspace{-12mm} G_{1\parallel}(t) = \frac{\chi_1}{\ep} e^{-t/\tau_{\rm L}},~~
 G_{2\parallel}(t) = 2\frac{1-\ef}{3\ep\ef}\chi_1e^{-t/\tau_{\rm L}}.
\ena 
where  the initial rapid 
decreases  of $G_{2\perp}(t)$ and $G_{2\parallel}(t)$  
are  $(\ef-1)/4\pi$ and $(1-1/\ef)/4\pi$, respectively. 
The FL transforms 
of $-d G_{i\perp} (t)/dt$ and  $-d G_{i\parallel}(t)/dt$ 
are  equal to 
the right-hand sides of Eqs.(90) and (91) 
and those of Eqs.(93) and (94), respectively. 

In statistical-mechanical  
theories\cite{Bagchi,Mukamel,Deutch,Mad,Lebe}, 
 inhomogeneous, oscillatory    fields  
were applied  fictitiously   without electrodes.  
These fields  are   transverse or longitudinal, 
  so   the  combined interaction energy is written  as     
\be 
{\cal H}^{\rm inh}_{\rm ext}= -\int\hspace{-1mm}  
 d{\bi r}~\Big[  {\rm Re}({\bi {\cal E}}_{T}) \cdot {\bi p}_\perp 
+ {\rm Re}({\bi{\cal E} }_L) \cdot {\bi p}_\parallel  \Big],   
\en 
where   ${\rm Re}(\cdots)$ denotes taking the real part.
Here,   $ {\bi{\cal E} }_T$ is a transverse radiation field, 
while    $ {\bi{\cal E} }_L$ is a longitudinal field 
  tending  to ${\bi E}_0$ in Eq.(85) in the homogeneous limit. 
Then, at wave number $q$ and frequency $\omega$, 
the linear response relations  for ${\bi p}_\perp$ 
and ${\bi p}_\parallel$   are   given by 
\be 
 \av{{\bi p}_\perp}= \chi^*(q,\omega){{\bi{\cal E}}}_T,~~
\av{{\bi p}_\parallel} 
={\alpha^*(q,\omega)}{\bi{\cal E}}_L,
\en 
where     $\chi^*(q,\omega)$ and ${\alpha^*(q,\omega)}$ 
are the FL transforms of 
$-\p G_{\perp}(q,t)/\p t$ and $-\p G_{\parallel}(q,t)/\p t$, 
respectively,  with  $ G_{\perp}(q,t)$ and $G_{\parallel}(q,t)$ 
being defined in Eq.(80). 
Then, the $(q,\omega)$-dependent  transverse and longitudinal 
dielectric constants can be 
defined by\cite{Bagchi,Mukamel,Deutch,Mad,Lebe}  
\bea
&&\ep_T(q,\omega)   = 
1+4\pi \chi^*(q,\omega),\\
&&\ep_L (q,\omega)  
= [1-4\pi {\alpha^*(q,\omega)}]^{-1}.
\ena 
For  large  $H$, these results  
 can be used for $q\gg \pi/H$ in the bulk.  
As   discussed   below Eq.(81), 
    $ \chi^*(q,\omega) \to \chi^*(\omega)$ 
and $ \alpha^*(q,\omega)\to \alpha^*(\omega)$ 
as   $q \to 0$, leading to  Eqs.(97) and (98).  
However,    we have derived  $\chi^*(\omega)$ in Eq.(97) 
 from  $\alpha^*(\omega)$ via Eq.(88) 
not using    the  transverse  fields.

 Chandra and Bagchi\cite{Bagchi} 
calculated $\ep_T(q,\omega)$ and $ \ep_L(q,\omega)$.   
Skaf {\it et al.}\cite{Lada} 
obtained  $ G_{\perp}(q,t)$ and  
$G_{\parallel}(q,t)$  for methanol, where  
 $ G_{\perp}(q,t)$ was much larger and  decayed 
much slower than   $G_{\parallel}(q,t)$ 
at the smallest wave number 
$0.24/{\rm \AA}$ in their simulation in accord with  Eqs.(78) and (79). 
Bopp {\it et al}.\cite{Bopp}  calculated 
$G_{p\parallel}(q,t)$ for water.

\subsection{Linear response. I: Total polarization}
In  the parallel plate geometry,  we can treat 
  the bare field ${\bi E}_0= E_0^* {\bi e}_z$ 
as the applied field controlling the electrode  charge 
$Q_0=L^2{\bar\sigma}_0$.  
Then,  its  conjugate variable is    $M^{\rm tot}_z$ in Eq.(51). 
 If $ E_0^*  \propto e^{i\omega t}$,  
we fix the amplitude $|E_0^*|$.  
We  present  the interaction energy in the form,   
\be 
{\cal H}^{\rm I}_{\rm ext}= -{\rm Re}( E_0^*)M^{\rm tot}_z 
=-4\pi{\rm Re}( {\bar\sigma}_0)M^{\rm tot}_z .
\en 
In our  scheme at fixed $Q_0$, 
the polarization fluctuations  obey the perturbed  distribution 
$(1- {\cal H}^{\rm I}_{\rm ext}/k_BT){\cal P}_0$, where 
${\cal P}_0\propto \exp(-{\cal F}/k_BT)$ 
is the distribution at $Q_0=0$.    Kubo himself derived  
 Eqs.(98)  and (107).

In   our theory,    Eqs.(30), (53), and (83) 
indicate that  the  relaxation functions of  $p_z$ and $p_{iz}$ 
in the bulk are     $G_{\parallel}(t)$    
in Eq.(79) and $G_{i\parallel}(t)$ in Eq.(100), respectively. Namely,     
\bea 
&&\hspace{-15mm}{\av{M^{\rm tot}_{z}(t) M^{\rm tot}_{z}(0)}}/{Vk_BT}
 =G_{\parallel}(t),
\nonumber\\
&& \hspace{-15mm}  
{\av{M^{\rm tot}_{iz}(t) M^{\rm tot}_{z}(0)}}/{Vk_BT} =G_{i\parallel}(t), 
\ena 
which   decay    with $\tau_{\rm L}$ from   Eqs.(79) and (102) 
with $G_{\parallel}(0)=\chi/\ep$ 
and $G_{i\parallel}(0)=\chi_i/\ep$. 
As in Eq.(98),  $\av{  p^*}/E_0^*$ and $\av{ p_i^*}/E_0^*$ 
are the FL transforms of 
$-dG_{\parallel}(t)/dt$ and  
$-dG_{i\parallel}(t)/dt$, respectively. 

The dielectric response is  heterogeneous  near 
a solid surface,  a two-phase interface,  a lipid bilayer,  
and so on,  
which was  calculated   in the static\cite{Hansen,Netz,Feller}  and  
 oscillatory cases\cite{Klapp,Netz1,Mon} with $\ef=1$. 
 In these papers, a polar fluid is 
 confined between   nonconducting  walls 
and the interaction energy is of the form 
${\cal H}_{\rm int}= - {\bi {\cal E}} \cdot {\bi M}^{\rm tot}$, 
where    $ {\bi {\cal E}} =({\cal E}_x, {\cal E}_y, {\cal E}_z)$ 
is a homogeneous   applied field. 
They  obtained     $z$-dependent dielectric constants,  
$\ep_\perp(z)$ and  $\ep_\parallel(z)$, where    
  the subscripts $\perp$ and $\parallel$ 
denote the directions  
 orthogonal  and  parallel to the surface, 
respectively.  For example, 
$1-1/\ep_\perp(z)=4\pi\av{p_z({\bi r})M_z}/k_BT$ 
at $\omega=Q_0=0$.

\subsection{Linear response. II: Embedded sphere}

Using $\chi_d$ in Eq.(25) and  
the time-correlation function 
$G_{\rm or}(t) $ for ${\bi p}_1$   in Eq.(76), we 
define   a frequency-dependent 
 orientational susceptibility by 
\bea 
&&\hspace{-1cm}\chi_d^*(\omega) 
= G_{\rm or}(0) -i\omega \int_0^\infty dt 
e^{-i\omega t} G_{\rm or}(t) 
 \nonumber\\
&&=\frac{\chi_d}{2+ \ef/\ep}
 \Big[ \frac{2}{1+ i\omega\tau_{\rm D}} 
+ \frac{\ef/\ep}{1+ i\omega\tau_{\rm L}}\Big]. 
\ena
Then,    use of  the Debye form of   
$\ep^*(\omega)$ in Eq.(89) yields  
  a frequency-dependent  KF equation,        
\be
\frac{3(\ep^*(\omega)-\ef)(2\ep^*(\omega)+\ef)}{4\pi\ep^*(\omega)(\ef+2)^2}
= \chi_d^*(\omega),
\en 
which  can be derived  with the aid of the relation,  
\be 
(\ep^*-\ef)\Big(2+\frac{ \ef}{\ep^*}\Big)
= \frac{\ep-\ef}{1+i\omega\tau_{\rm D}} 
\Big[2+ \frac{\ef}{\ep}\cdot
\frac{1+ i\omega\tau_{\rm D}}{1+i\omega\tau_{\rm L}}\Big].\nonumber 
\en   

To understand Eq.(110) in the linear response theory, 
we introduce   the frequency-dependent 
directing field, 
\be 
E_d^*= \frac{\ep^*(\omega)(\ef+2)}{2\ep^*(\omega)+\ef}E^*.
\en 
which  gives   Eq.(23) as $\omega\to 0$. As in Eq.(84), 
we have   ${\bi E}= E^*{\bi e}_z$ and 
${\bi E}_d= E_d^*{\bi e}_z$ in the parallel plate geometry.  
Using   $ \chi_1^*(\omega) $  in Eq.(90) and 
  $ \chi_d^*(\omega) $ in Eq.(109) we find 
\be 
p_1^*= \chi_1^*(\omega)  E^*=  \chi_d^*(\omega)E_d^*, 
\en 
which tends to  Eqs.(22) and (50) as $\omega \to 0$.

 If we suppose a mesoscopic  sphere  with volume 
 $v=4\pi R^3/3\ll V$ 
in the bulk,    ${\bi E}_d= E_d^*{\bi e}_z$ is 
 the  externally applied, oscillating   field,  whose conjugate 
variable is  the interior ${M}_{1z}$ in Eq.(45).
Thus, we propose the  effective   interaction energy  
in the sphere, 
\be 
{\cal H}^{\rm II}_{\rm ext}
=  -{\rm Re}( {\bi E}_d) \cdot {\bi M}_1 .
=  -{\rm Re}( {E}^*_d) { M}_{1z} .
\en 
Then, the relaxation function of  ${ p}_{1z}$ 
is   given by   $G_{\rm or}(t)$ in Eq.(76). 
From  Eqs.(109) and (112) we confirm   
\be 
G_{\rm or}(t)=
 \frac{\av{M_{1z}(t)M_{1z}(0)}}{vk_BT}
=\frac{\av{{\bi M}_{1}(t)\cdot {\bi M}_{1}(0)}}{3vk_BT}.
\en 
As discussed  below Eq.(50), we cannot  
 calculate the linear response of 
${\bi p}_2$ from Eq.(113).

Since Eq.(110)  is a natural generalization of the 
KF equation, it   
has been  presented by many 
authors (mostly for $\ef=1$)
\cite{Neu,Mukamel,Deutch,Ful2,Gla,Ni,Wi,Cole,Zw,Mason}, 
but  the  relaxation  function 
$G_{\rm or}(t)$ has not been 
calculated explicitly.  
Nee and Zwanzig\cite{Zw}  
 assumed $ {\bi p}_2={\bi p}-{\bi p}_1=
 [(\ef-1)/4\pi]{\bi E}$, 
 the effective dipole moment 
$\mu= \mu_0 (\ef+2)/3$, and a modified  
directing field ($=[3/(\ef+2)]{\bi E}_d)$ to obtain Eq.(113), 
following Fr$\ddot{\rm o}$hlich\cite{Fro}   
(see the last paragraph in Sec.IIIB).

\subsection{Linear response. III: Embedded sphere}

To  generalize Eq.(49) to a frequency-dependent one,    
we introduce  the frequency-dependent cavity field,     
 \be 
E_c^* = \frac{3\ep^*(\omega)}{2\ep^*(\omega)+1}E^*,     
\en 
which gives  ${\bi E}_c$ in Eq.(50) as $\omega\to 0$. 
Again we suppose  a mesoscopic sphere in the bulk.
Then,
  ${\bi E}_c= E_c^* {\bi e}_z $ is the oscillating applied field,  
whose   conjugate variable is   the interior ${\bi M}$ in Eq.(45). We 
find   the interaction energy,  
\be 
{\cal H}^{\rm III}_{\rm ext}
= -{\rm Re}({\bi E}_c) \cdot {\bi M}=
 -{\rm Re}({ E}^*_c)  { M}_z .
\en 

In the situation (80), use of Eqs.(97) and (98) gives 
\be 
p^*\hspace{-1mm}
= \chi^*(\omega){E^*}\hspace{-0.4mm}
 = \alpha^*(\omega){E_0^*} 
= [\frac{2\chi^*(\omega)}{3}+\frac{\alpha^*(\omega)}{3}]E_c^* .
\en 
Thus, $p^*/E_c^*$  is equal to 
 the FL transform of $-dG(t)/dt$,
 where $G(t)= 2G_\perp(t)/3+G_\parallel(t)/3$ 
is given   in Eq.(77). Using  Eqs.(89) and (92) 
we then obtain  
\bea 
&&\hspace{-15mm} (\ep^*(\omega)-1) 
\frac{2\ep^*(\omega)+1}{12\pi\ep^*(\omega)}
=G(0) -i\omega \int_0^\infty \hspace{-2mm}
dt e^{-i\omega t} G(t) 
\nonumber\\
&&\hspace{-10mm} = G_{\rm f}(0)  + 
\frac{\ep-\ef}{12\pi} \Big[ \frac{2}{1+ i\omega\tau_{\rm D}}+
\frac{1/\ep\ef}{1+ i\omega\tau_{\rm L}} \Big],   
\ena 
where $G_{\rm f}(0)= (\ef-1)(2+1/\ef)/{12\pi}$. 
This equation    coincides with Eq.(110) for $\ef=1$. 
We also confirm 
\be
G(t)= \frac{\av{{M}_z(t)\cdot { M}_z(0)}}{vk_BT}
= \frac{\av{{\bi M}(t)\cdot {\bi M}(0)}}{3vk_BT}.
\en 
 Fulton\cite{Ful2} derived   the first line of Eq.(118) 
for $\ef=1$.

The relaxation functions of  ${\bi p}_i $  are given by 
\be 
  G_{i}(t) = \frac{2}{3}G_{i\perp}(t)+\frac{1}{3} G_{i\parallel}(t)
 =\frac{ \av{{\bi M}_i(t)\cdot{\bi M}(0)}}{3vk_BT},   
\en 
where $G_{i\perp}(t)$ and $ G_{i\parallel}(t)$ are 
defined in  Eqs.(99) and (100). 
As in Eq.(117) $p_i^*/E_c^*=
 [{2}\chi_i^*(\omega)+\alpha_i^*(\omega)]/3 $ 
 are equal to   the FL transforms of $- d G_{i}(t) /dt$.

\section{Fluctuations and linear response 
in  fixed-potential condition}
 In 
simulations\cite{Wang,La,Le,Limmer,T2,Takae,T1,Hau,P1,P3,P4,P5,P6,P7,Sprik1}, 
 the applied electric field $E_a= \Phi_a/H$ in Eq.(17) can 
 be fixed  between two metal plates. 
This is a typical experimental method, 
where the  condition    $Q_H= -Q_0$ 
at fixed  $E_a$ is   realized  
by attachment of   
 an external circuit  to the two metal electrodes. 
We assume  this  fixed-potential condition along the $z$ axis, 
imposing  the periodic boundary condition along the 
$x$ and $y$ axes.  In the oscillatory case, 
 the amplitude $|E_a|$ is fixed.   
There are no  electric charges in the fluid. 

\subsection{Interaction energy at fixed $E_a$}
 
At fixed $E_a$,  the mean surface charge density  
${\bar\sigma}_0=Q_0/L^2$ 
fluctuates. Since the lateral average of 
$D_z$ is independent of $z$ from $\nabla\cdot{\bi D}=0$, we find   
\be 
4\pi{\bar \sigma}_0 = {\bar E}(z)+4\pi {\bar p}(z) 
= {\bar{\bar E}}+4\pi {\bar{\bar p}}, 
\en  
which holds at any $z$.
 Here,  ${\bar E}(z)$ and ${\bar p}(z)$ 
are the lateral averages of the $z$ components 
 $E_z({\bi r})$ and $p_z({\bi r})
=p_{1z}({\bi r})+p_{2z}({\bi r})$, 
while  $\bar{\bar E}$ and $\bar{\bar p}$ are their 
 cell averages. Namely,   
\bea 
&&\hspace{-11mm}
{\bar E}(z)=\frac{1}{L^2} \int 
 d{\bi r}_\perp E_z({\bi r}), 
~~{\bar p}(z)=\frac{1}{L^2} \int d{\bi r}_\perp p_z({\bi r}),
\nonumber\\
&&\hspace{-11mm}
{\bar{\bar E}}=
 \frac{1}{H}\int\hspace{-1mm} dz~ 
{\bar E}(z), ~~{\bar{\bar p}}=\frac{1}{H}  
\int \hspace{-1mm}dz~ {\bar p}(z)=\frac{1}{V}M_z^{\rm tot},
\ena 
where 
$\int \hspace{-0.5mm} d{\bi r}_\perp$ and $\int dz$ 
denote  the integrations with respect to  ${\bi r}_\perp=(x,y)$     
and  $z$, respectively,  in the cell. 
To avoid confusion, the subscript $z$  for  the $z$ components   is   not   
written for  these space averages. 
The lateral and cell averages  consist of the Fourier components 
with $q_x=q_y=0$ and  ${\bi q}={\bi 0}$, respectively. 
The correlations  with   ${\bi q}\neq {\bi 0}$
are insensitive to the boundary condition. 

Including the surface effect we rewrite  Eq.(17) as   
\be 
E_a = {\bar {\bar E}}  +{{\bar \sigma}_0}/CH.
\en   										The statistical averages 
of  ${\bar{\bar E}}$,  ${\bar{\bar p}}$, 
and ${\bar \sigma}_0$ are given by 
\be 
 \av{{\bar{\bar E}}}_a=E_b, ~~ \av{{\bar{\bar p}}}_a= \chi E_b, 
~~\av{{\bar \sigma}_0}_a=\ep E_b/4\pi,
\en 
in terms of $E_b$ in Eq.(18).  Hereafter,   
$\av{\cdots}_a$ denotes taking   the statistical   
average at fixed $E_a$.
From Eqs.(121) and (123) their fluctuation parts 
 $\delta{\bar{\bar E}}={\bar{\bar E}}- E_b$, 
 $\delta{\bar{\bar p}}={\bar{\bar p}}-\chi E_b$, and 
 $\delta{\bar \sigma}_0={\bar \sigma}_0-\ep E_b/4\pi$ 
are related by 
\bea 
&&\delta{\bar \sigma}_0= \delta{\bar{\bar p}}/(1+1/4\pi CH)
\cong  \delta{\bar{\bar p}},\nonumber\\ 
&&\delta{\bar{\bar E}}= -\delta{\bar \sigma}_0/CH,
\cong -\delta{\bar{\bar p}}/CH,~~~ 
\ena
where     $1/4\pi C=\ell_{\rm w}/\ep\ll H$ from Eq.(19) 
for $\ep\gg 1$.

For stationary $E_a$, 
the polarizations obey the  distribution 
$\propto \exp[ - {\tilde{\cal F}}/k_BT]$ with  
   ${\tilde{\cal F}}$   in Eq.(15). 
This indicates that 
 the interaction energy  is given  by   
\be
{\cal H}^{\rm IV}_{\rm ext}= -{\rm Re}(\Phi_a) {Q_0}
=-{\rm Re}({ E}_a) {M}^{\rm tot}_z , 
\en 
where  the fluctuation parts of 
 $Q_0$  and $L^2{\bar{\bar p}}={M}^{\rm tot}_z/H$ 
coincide for $d\ll H$ from Eq.(125). 
In our scheme,  the polarization fluctuations 
obey the  perturbed distribution 
${\cal P}_a= [1- {\cal H}^{\rm IV}_{\rm ext}/k_BT] {\cal P}_a^0$ 
in the linear order,  
where ${\cal P}_a^0\propto \exp(-{\cal F}/k_BT)$ 
is the distribution at  $E_a=0$ under   the  global 
electrostatic constraints (121) and (123). 
We use  Eqs.(121)-(126)  even 
in the oscillatory case.

\subsection{Nonlocal correlations for stationary  $E_a$}

Because the problem is multifold,   we first assume   
 stationary $E_a$ and  neglect  the surface effect. 
Then,   ${\bar{\bar E}} =E_a=E_b$ and ${\bar E}(z) =E_a-4\pi 
({\bar p}(z)- {\bar{\bar p}})$.

In $\tilde{\cal F}={\cal F}-H E_a{\bar\sigma}_0$  we 
 pick up the Fourier  components of $ p_z({\bi r})$  
with $q_x=q_y=0$. We then   obtain 
   the one-dimensional   free energy density 
${\cal F}_{\rm 1D}= {\tilde{\cal F}}/L^2$ as 
\bea 
&&\hspace{-12mm} {{\cal F}_{\rm 1D}}=
 \int \hspace{-1.6mm}
dz\Big[ \frac{1}{8\pi} {\bar E}(z)^2 +
\frac{1}{2\chi} {\bar{p}(z)}^2 \Big] 
-H E_a{\bar\sigma}_0 \nonumber\\
&&\hspace{-8mm} 
= \int \hspace{-1.6mm}
dz \frac{ \ep}{2\chi}
\Big[\delta{\bar p}(z)-\delta {\bar{\bar p}} \Big]^2
+\frac{H}{2\chi}(\delta {\bar{\bar p}})^2  
-\frac{H\ep}{8\pi}E_a^2,
\ena 
where  $ \delta{\bar{p}}(z)={\bar{ p}}(z)-\chi E_a$, 
  The last term in the second line is 
 the minimum of ${\cal F}_{\rm 1D}$. 
The  variable  $\bi s$   in Eq.(29)  
  is  decoupled from  $\bi p$, so    
 its contribution is not written here.  

The fluctuation part  of ${\cal F}_{\rm 1D}$  
is written in the double integral form    $\int dz \int dz' c(z,z')
\delta{\bar p}(z) \delta{\bar p}(z')/2$ with 
\be  
c(z,z')=(\ep/\chi)\delta(z-z') - 4\pi /H.
\en 
Then, we find  the  one-dimensional   correlation function,        
\bea 
&&\hspace{-14mm} {G}_{\rm 1D}(z,z')= 
 \av{\delta{\bar p}(z) \delta{\bar p}(z')}_a L^2/k_BT 
\nonumber\\
&&= ({\chi}/{\ep})\delta(z-z')+ 4\pi {\chi}^2/{\ep H},   
\ena 
where we use  $\int dz'' c(z,z'')G_{\rm 1D}(z'',z')=\delta(z-z')$.  
From Eq.(125)  we obtain the variance relations     at fixed $E_a$, 
\bea 
&& \hspace{-13mm}
V\av{(\delta {\bar \sigma}_0)^2}_a=
V\av{(\delta {\bar{\bar p}})^2}_a 
= \frac{\av{(\delta M_z^{\rm tot})^2}_a}{V}={k_BT \chi}, 
\ena 
Here, all $\av{\delta M_{i\alpha}^{\rm tot}\delta M_{j\beta}^{\rm tot}}$ 
are given by   Eq.(56), but 
only the $zz$ components   $G^{ij}_{zz}({\bi r})$  
  acquire  nonlocal parts given by 
$4\pi \chi_i\chi_j/\ep V$ for  $H\ll L$. 
 
In contrast,   at fixed $Q_0$, 
the relation  $\delta{\bar E}=
-4\pi \delta{\bar{p}}$ gives 
  $G_{\rm 1D}(x,x')=  (\chi/\ep) \delta(z-z')$ 
with  no  nonlocal correlation in  the $z$ direction.
Furthermore, since Eq.(53) holds, 
we find no nonlocal correlation also in the $xy$ plane,

\subsection{ Surface effect   for stationary  $E_a$}

Next, to include  the surface effect 
for stationary $E_a$, 
we   add  a surface free energy density  
 to   ${{\cal F}_{\rm 1D}}$ in Eq.(127): 
 \bea 
&&\hspace{-10mm} {\cal F}_{\rm 1D}^{\rm tot} 
= {{\cal F}_{\rm 1D}}+ {{\bar \sigma}_0}^2/2C 
= \int \hspace{-1.6mm}
dz \frac{ \ep}{2\chi}
\Big[\delta{\bar p}(z)-\delta {\bar{\bar p}} \Big]^2
\nonumber\\
&&\hspace{-0mm}
 + (H+\ell_{\rm w}){(\delta {\bar{\bar p}})^2}/2\chi   
 -{H\ep }E_aE_b/8\pi.
\ena 
where  $ \delta{\bar{p}}(z)={\bar{ p}}(z)-\chi E_b$    
and   $ \delta{\bar{\bar{p}}}=\delta{\bar{{\sigma}_0}}$.  
The last term  is equal to the thermodynamic  
free energy $-\Phi_a \av{{Q}_0}_a/2\pi$ 
 divided by $L^2$.  
See Appendix A for 
 a derivation of the  surface free energy 
density ${{\bar \sigma}_0}^2/2C $. 

With Eq.(131)  $G_{\rm 1D}(x,x')$ in Eq.(129) is changed to      
 \be 
G_{\rm 1D}(z.z')= \frac{\chi}{\ep}\Big[\delta(z-z')- \frac{1}{H}\Big]+ 
\frac{\chi}{ H+\ell_{\rm w}}. 
\en 
Then,  we obtain   the variance relations, 
\bea 
&&\hspace{-12mm}
 V \av{(\delta{\bar{\sigma}}_0)^2}_a=
  V\av{(\delta{\bar{\bar p}})^2}_a
={\av{(\delta M_z^{\rm tot})^2}_a}/{V}
\nonumber\\
&&={k_BT \chi}/{(1+ \ell_{\rm w}/H)}.
\ena 
The typical size of $\delta{\bar \sigma}_0$ 
is small and is estimated as 
\be 
|\delta{\bar \sigma}_0| 
\sim 0.3 e/(L\sqrt{\ell_B (H+\ell_{\rm w})}),  
\nonumber 
\en 
where  $\ell_B$  is  the Bjerrum length.  
From Eq.(30) we find   
\be 
{\av{\delta M^{\rm tot}_{i z} 
\delta M^{\rm tot}_{j z}}}_a =V {k_BT}\Big[
\chi_\perp^{ij} - \frac{\chi_i\chi_j}{\chi(H+ \ell_{\rm w})}\ell_{\rm w}\Big].  \en 
These  variances among the $z$ components   are 
very different from  those  
in Eqs.(53) at $Q_0=0$, while those  among  
the $x$ and $y$  components 
are commonly given by Eq.(53).  
Furthermore, let  $G^a_{\alpha\beta}({\bi r})
= \av{p_\alpha({\bi r})p_\beta({\bi 0})}_a/k_BT$   be 
the  space-correlation function  of 
$\bi p$ at  $E_a=0$, which  differs from 
 $G_{\alpha\beta}({\bi r})$    in Eq.(44)  only for $\alpha=\beta=z$ as\cite{Takae}    
\be 
{ G}^a_{zz}({\bi r}) 
={ G}_{zz}({\bi r})-\chi/\ep V+  \chi/{ [V(1+ \ell_{\rm w}/H)]} . 
\en

Takae and the present author\cite{Takae} performed a 
 fixed-potential  simulation. 
With  $\ell_{\rm w}=2H$, 
they found   Eq.(133) for the variance of 
$M_z^{\rm tot}$, 
  Eq.(53) or Eq.(56) for those of 
 $M_x^{\rm tot}$ and $M_y^{\rm tot}$,   
 and Eq.(135) for ${ G}^a_{zz}({\bi r})$.

\subsection{Surface charge fluctuations}

The local surface charge densities 
$\sigma_0$ and $\sigma_H$  
are  equal to  $ \pm D_z/4\pi$ at $z=0$ and  $H$ in 
the continuum electrostatics. In some 
simulations\cite{P1,P3,Hau,T1,T2,Takae}, 
this relation was used   on  smooth surfaces. 
  Siepmann and Sprik\cite{Sprik2}  presented an electrode model, where 
atomic particles in the electrodes 
have charges varying continuously 
to realize the metallic boundary condition. 
Here, we divide $\sigma_0$ and $\sigma_H$  
into  the lateral averages 
${\bar{\sigma}}_0 $ and ${\bar{\sigma}}_H(=-{\bar{\sigma}}_0) $ 
and the  inhomogeneous  parts    
$\sigma_0^{\rm inh}
=\sigma_0-{\bar{\sigma}}_0 $ and 
$\sigma_H^{\rm inh}=\sigma_H-{\bar{\sigma}}_H$, where 
${\bar{\sigma}}_0 $  is coupled to $M^{\rm tot}_z$ as in Eq.(125).  

We   consider the in-plane correlation 
 of the  deviation  $\delta\sigma_0({\bi r}_\perp)
=\sigma_0({\bi r}_\perp) -\ep E_b/4\pi$ at $z=0$. 
It behaves as  
\bea 
&&\hspace{-10mm}
 {\cal G}_0({\bi r}_\perp-{\bi r}_\perp') =
\av{\delta\sigma_0({\bi r}_\perp)\delta\sigma_0({\bi {\bi r}_\perp'})}_a/k_BT
\nonumber\\ 
&&\hspace{-8mm} 
=S_0  \delta ({\bi r}_\perp-{\bi r}_\perp')- S_0/{L^2} +
\chi/{ [L^2(H+ \ell_{\rm w})]} , 
\ena 
where $\delta ({\bi r}_\perp)$ is a localized function 
 with $\int d{\bi r}_\perp \delta ({\bi r}_\perp)=1$.   
The last two terms are nonlocal.
We set $S_0= \int d{\bi r}_\perp{\cal G}_0({\bi r}_\perp)
$ in the limit $L\to\infty$  (see Appendix A). 
Then, the  in-plane integral of ${\cal G}_0({\bi r}_\perp)$   becomes   
\be 
\int \hspace{-1mm}d{\bi r}_\perp 
{\cal G}_0({\bi r}_\perp)=\frac{ \av{(\delta {Q}_0)^2}_a}{L^2k_BT}
=\frac{\chi}{H+\ell_{\rm w}}, 
\en 
where $\delta Q_0=L^2\delta{\bar\sigma}_0$. 
This is in accord   with  Eq.(133).  
However, both at fixed $Q_0$ and at fixed  $E_a$,  the  
in-plane correlation of 
$\sigma_0^{\rm inh}=\sigma_0-{\bar{\sigma}}_0 $  is given by   
\bea 
&&\hspace{-15mm}
\av{\sigma_0^{\rm inh}({\bi r}_\perp)\sigma_0^{\rm inh}
({\bi {\bi r}_\perp'})}=
\av{\sigma_0^{\rm inh}({\bi r}_\perp)\sigma_0^{\rm inh}
({\bi {\bi r}_\perp'})}_a\nonumber\\
&&\hspace{-6mm} = k_BT[S_0  \delta ({\bi r}_\perp-{\bi r}_\perp')
- S_0/{L^2}],
\ena  
which follows from Eq.(136) at fixed $E_a$ 
and is suggested  by   the last paragraph in Sec.IV at fixed  $Q_{0}$.

In our   simulation\cite{T2}, the correlation length $\xi_s$ of 
$ {\cal G}_0({\bi r}_\perp)$ was of order $1{\rm \AA}$ 
(for $|{\bi r}_\perp|\ll L$) and  $S_0 $ was about $0.3/$nm.
Thus,  in Eq.(136), the third term is 
comparable to the second one 
unless $H\gg \ell_{\rm w}$. 
We also found that   the electric fields  produced by 
 $\sigma_0^{\rm inh}$ and 
 $\sigma_H^{\rm inh}$ in the fluid 
decay rapidly outside the Stern layers. 
On the other hand,  in the previous papers on the surface 
charges\cite{Limmer,La,P7}, use has been made of 
   $\av{(\delta {Q}_0)^2}_a/L^2k_BT = S_0=C_0$ 
without   the nonlocal terms in Eq.(136), where 
the global conditions (121) and (123) were not 
accounted for.

\subsection{Dynamics for total polarizations }
 
We  examine  dynamics for   $E_a \propto e^{i\omega t}$ 
with $\omega \tau_{\rm f}\ll 1$ including   the surface effect. 
We first assume $\omega\tau_s\ll 1$, where 
$\tau_s$ is   the relaxation time  of 
 the surface polarization (see Appendix A). 
Then,   the surface potential drop 
is  given by  ${\bar \sigma}_0/C$   
and     Eqs.(121) and (123) give  
\be 
{\bar E}(z,t) = E_a(t) -{\bar\sigma}_0(t)/CH  
-4\pi[{\bar p}(z,t)-{\bar{\bar p}}(t)] , 
\en 
where  ${\bar \sigma}_0\cong {\bar{\bar p}}+E_b/4\pi$ 
from Eq.(125). 
Hereafter, we suppress  the space-time dependence 
of the variables if no confusion occurs. 
 For  $\omega\tau_s\gs 1$ we 
need to replace $\ell_{\rm w}$ by its frequent-dependent 
generalization (145) below.

We take the lateral and cell 
averages of the $z$ component of Eq.(67) 
writing those of  $p_{1z}$ as  ${\bar p}_1$ and $\bar{\bar p}_1$,   
where   the lateral  average of $E_{0z}$ is $4\pi{\bar \sigma}_0$. 
 Setting    ${\bi \xi}={\bi 0}$,  
 we   find   
\be 
{\bar p}- {\bar{\bar p}}= 
\frac{\ef+2}{3\ef}({\bar p}_1- {\bar{\bar p}}_1),~~
{\bar{\bar p}}=\frac{\ef-1}{4\pi}{\bar{\bar E}}
+ \frac{\ef+2}{3}{\bar{\bar p}}_1 . 
\nonumber
\en 
Using these relations  and assuming  $\ep\gg \ef$ we have 
\be 
\chi_1 {\bar E}\cong \chi_1 E_a- \ell_{\rm w}{\bar{\bar p}}_1/H 
-(\ep/\ef-1)({\bar p}_1-{\bar{\bar p}}_1), 
\en   
where the second term is amplified by $\chi_1$. 
Then,  the  lateral average of the $z$ component of   Eq.(73) gives  
\bea
&&\hspace{-12mm} \tau_{\rm D} \frac{\p}{\p t}{\bar p}_{1}
=\chi_1 E_a- (1+{\ell_{\rm w}}/{H} )  {\bar{\bar p}}_{1} 
- \frac{\ep}{\ef}( {{\bar p}}_{1}-{\bar{\bar{p}}}_{1}),
\ena 
which   gives ${\bar p}_1= {\bar{\bar p}}_{1} =\chi_1E_b$ 
in equilibrium at $\omega=0$. 

From Eq.(141) the homogeneous variables 
  ${\bar{\bar{p}}}_1$,  ${\bar{\bar{p}}}$, 
and  ${\bar{\sigma}}_0$   
decay as $\exp(-t/\tau_{\rm D}')$ with a modified 
relaxation time, 
\be 
\tau_{\rm D}'= \tau_{\rm D}/(1+\ell_{\rm w}/H)=\tau_{\rm L}\ep_{\rm eff}/\ef.  
\en 
However, the inhomogeneous part 
  ${{\bar p}}_{1}-{\bar{\bar{p}}}_{1}$  decays with $\tau_{\rm L}$ 
as in the fixed charge condition.  
For example, at $E_a=0$, the time-correlation of 
$M_{1z}^{\rm tot}(t)= V{\bar{\bar p}}_1(t)$ 
decays as 
\be 
\av{M_{1z}^{\rm tot}(t) M_{1z}^{\rm tot}(0)}/Vk_BT 
=   e^{-t/\tau_{\rm D}'}D_{11}  ,
\en
where $D_{11}=\chi_\perp^{11} - 
{\chi_1^2\ell_{\rm w}}/[{\chi(H+ \ell_{\rm w})}]$ 
from Eq.(134). This equation  is  similar 
to that in Eq.(84) for the periodic boundary condition. 
We generalize  Eq.(76) to  
\bea 
&& \hspace{-10mm} \av{{{\bi p}_1} ({\bi r},t)\cdot
{{\bi p}}_1({\bi r}',0)}/{k_BT} =3
G_{\rm or}(t)\delta({\bi r}-{\bi r}') \nonumber\\
&&  -e^{-t/\tau_{\rm L}} \chi_\perp^{11}\ef/\ep V
+e^{-t/\tau_{\rm D}'} D_{11}/{V}.
\ena

In  Appendix A, the Stern layers will  be 
approximated as  thin  films 
with a  thickness $d$ and a  dielectric constant 
$\ep_{s} (\ll \ep)$. 
  Using the surface relaxation time $\tau_s$ 
we will obtain the frequency-dependent 
surface electric length,  
\be 
\ell_{\rm w}^*(\omega) 
= \ell_{\rm w} (1+i\omega \tau_s\ep_s)/({1+i\omega \tau_s}),
\en   
where  $\ell_{\rm w}=2d\ep/\ep_s$ is the static limit. 
 For  $\omega\tau_s\gg 1$, we have 
$\ell_{\rm w}^*(\omega)\to 2d\ep$ (the value for empty  
films)\cite{P6,P3}.

\subsection{Linear response  at fixed $E_a$ }

We  examine  the linear response for $E_a \propto e^{i\omega t}$.
We use the static length $\ell_{\rm w}$  assuming  
  $t\gg \tau_s$ in the time correlations 
and  $\omega\tau_s\ll 1$ in the frequency-dependent relations. 
For $\tau_s^{-1}\ls \omega \ll \tau_{\rm f}^{-1}$, 
we should use   $\ell_{\rm w}^*(\omega)$ in Eq.(145).

We consider the  relaxation function of   $p_z$ 
at $E_a=0$, written as $K_a(t)$. 
From Eq.(125)  and (126) 
it is expressed in terms of
 the following correlations at $E_a=0$:       
\be 
K_a(t)=\frac{\av{M_z^{\rm tot}(t)M_z^{\rm tot}(0)}_a}{Vk_BT}
={V}
\frac{\av{{\bar\sigma}_0(t){\bar\sigma}_0(0)}_a }{k_BT}. 
\en 
We can then relate $K_a(t)$ to $G_\perp(t)$  in Eqs.(78) as 
\be
K_a(t)= G_\perp(t' )/(1+\ell_{\rm w}/H), 
\en 
where  $t'=  t(1+\ell_{\rm w}/H)$ and  $t'/\tau_D=
t/\tau_D'$, so  $K_a(t)$  decays  with $\tau_{\rm D}'$.  
Notice that  $ G_\perp(t )$ also appears in Eqs.(97), 
where the surface potential drop is irrelevant at fixed $Q_0$.  
However,  $K_{a}(t)$ much  differs  from 
  the  relaxation function $G_\parallel(t)$ at $Q_0=0$  in Eq.(108), 
though they are both  expressed in the form 
of the  time correlation of $M_z^{\rm tot}$. 
  In the same manner,   using  $G_{i\perp}(t )$  in Eq.(99), 
the  relaxation function of   $p_{iz}$ are 
written as 
\bea 
&&\hspace{-13mm}
K_{ia}(t)=\frac{\av{M_{iz}^{\rm tot}(t)M_z^{\rm tot}(0)}_a}{Vk_BT}
= \frac{G_{i\perp}(t' )}{1+\ell_{\rm w}/H},  
\ena 
which are very different from  $G_{i\parallel }(t) $ in Eq.(108).

The frequency-dependent 
susceptibilities are defined by 
 $\chi_{a}^*(\omega)= p^*/E_a$  and 
 $\chi_{ia}^*(\omega)= p_i^*/E_a$, which are  the 
FL transforms of 
$- d K_a(t)/dt$ and $- d K_{ai}(t)/dt$, respectively. 
From Eqs.(78) and (101) they are written as   
\bea 
&&\hspace{-10mm}\chi^*_{a}(\omega) = 
\frac{1}{4\pi}\Big[\frac{{\ef-1}}{1+ \ell_{\rm w}/H}
+\frac{\ep-\ef}{1+ \ell_{\rm w}/H+i\omega\tau_{\rm D}}
\Big],\\
&&\hspace{-10mm}\chi_{1a}^*(\omega)
=\frac{ \chi_1}{1+ \ell_{\rm w}/H+ i\omega \tau_{\rm D}},\\ 
&&\hspace{-10mm}\chi_{2a}^*(\omega)
= \frac{\ef-1}{4\pi}\Big[\frac{1}{1+ \ell_{\rm w}/H}
+\frac{4\pi}{3}  \chi_{1a}^*(\omega)\Big],
\ena 
which  also follow directly from Eqs.(9) and (140). 
In large cells with   $H\gg \ell_{\rm w}$,    these susceptibilities  
  tend to $\chi^*(\omega)$ and $\chi^*_{i}(\omega)$ in  Eqs.(89)-(91), 

At $E_a=0$ we also calculate  the space-time correlation  
${ G}^{a}_{\alpha\beta}({\bi r},t) 
=\av{p_{\alpha}({\bi r},t) p_{\beta}({\bi 0},0) }_a/k_BT$. 
It   differs from $ {G}_{\alpha\beta}({\bi r},t)$  
 in Eq.(82) only for $\alpha=\beta=z$ as  
\be
{ G}^{a}_{zz}({\bi r},t) 
 = { G}_{zz}({\bi r},t) - [G_\parallel(t)-K_a(t)] /{V}. 
\en

In the bulk region,  the average  electric field  
is  $E_b$ in Eq.(18)   and   
the average electric induction 
is  $E_b+ 4\pi \chi^*_{a}(\omega)E_a=4\pi{\bar\sigma}_0$. 
Then,   we can   define 
the frequency-dependent dielectric   constant in the Debye form, 
\be
\ep^*_{b}(\omega) = 4\pi {\bar\sigma}_0/E_b
= \ef +({\ep-\ef})/({1+i\omega\tau_{\rm D}'}),
\en 
where  the time constant is  $\tau_{\rm D}'$ in Eq.(142). 
On the other hand, in the capacitor experiments, 
 the effective dielectric constant 
should be determined by   
\be 
\ep^*_{\rm eff}(\omega) = 4\pi {\bar\sigma}_0/E_a 
= \ep_b^*(\omega)/(1+ \ell_{\rm w}/H), 
\en 
which tends to $\ep_{\rm eff}$ in Eq.(2) as  $\omega\to 0$. 

\section{Summary and remarks} 

We have studied   statics and dynamics  of  dielectric fluids 
  in the linear regime in the continuum theory.
We  have  assumed 
 the fixed-charge and fixed-potential 
conditions  in a $L\times L\times H$ cell with $H\ll L$. 
We have also revealed consequences for the 
total polarizations in other boundary conditions. 
We have found nonlocal polarization correlations, 
which are essential in electrostatic systems. 
 Our main results are summarized below.\\ 
\indent 
(i) In Sec.IIA, we have started with Felderhof's  
dielectric free energy density $f $ in Eq.(3)\cite{Felder2}. 
It contains  the Lorentz term $\propto {\bi p}_1\cdot{\bi p}_2$ 
between the orientational  polarization ${\bi p}_1$ and 
the   induced one ${\bi p}_2$, which yields  the Lorentz internal 
field ${\bi F}$ in Eq.(8) governing ${\bi p}_2$ as in Eq.(9). 
From this $f$  we have obtained the susceptibilities $\chi_1$ and 
$\chi_2$   in Eq.(13), which   also follow 
from Onsager's theory\cite{Onsager}. 
In  Sec,IIB, we have examined  the effect of the 
 potential drop in the Stern layers to find  the 
effective dielectric constant $\ep_{\rm eff}$ in Eq.(2).
In Sec.IIC, we have introduced  the directing field 
${\bi E}_d$ governing ${\bi p}_1$ in Eqs.(21)-(24) 
and  the orientational susceptibility $\chi_d$ 
by ${\bi p}_1= \chi_d {\bi E}_d$ to reproduce 
 Onsager's results.

(ii) In Sec.III, we have calculated 
the static polarization correlations 
in the wave number range $\pi/H<q<\pi/a_m$ 
or in the distance range $a_m<r<H$, 
where $a_m$ is the molecular length. 
These fluctuations are insensitive to the boundary condition.  
 We have then 
obtained the Kirkwood-Fr$\ddot{\rm{o}}$hlich equation (25).

(iii) In Sec.IV, we have calculated the 
variances of the total polarizations ${\bi M}_i^{\rm tot}$. 
They are given by  Eq(53)  at  fixed   $Q_0$ and   
 by Eq.(56) in   the periodic boundary condition. 
We have also examined   Sprik's case\cite{Sprik,Sprik1,Cox}  
with    $\int_V d{\bi r}{\bi D}={\bi 0}$.
The nonlocal polarization correlations  
have also been examined. Their effects on the sphere integral 
${\bi M}_1= \int_{r<R} d{\bi r}{\bi p}_1$  
has been shown  in Eq.(58).

(iv) In Sec.V, we have examined the solvation  
free energy for given ${\bi p}_1$ and 
bare electric field ${\bi E}_0$, 
since   the Lorentz term is missing  in  Marcus'  free energy\cite{Marcus}.

(v) In Sec.VI, we have studied  polarization dynamics, 
where three relaxation times $\tau_{\rm D}$, 
$\tau_{\rm L}=\tau_{\rm D} \ef/\ep$, and $\tau_{\rm f}$ appear 
with $\tau_{\rm D}>\tau_{\rm L}\gg \tau_{\rm f}$. 
Then, the   polarization      time-correlation functions 
have been calculated.

(vi) In Sec.VII,     
we have given frequency-dependent 
linear response relations. 
The  relaxation functions\cite{Kubo} 
have been calculated  in analytic forms 
 in typical  situations.

(vii) In Sec.VIII, 
 we have controlled  
the applied field  $E_a= \Phi_a/H$ 
under the global constraints (121) and (123). 
Nonlocal   polarization 
correlations are produced by   surface-charge fluctuations.  
The variance and the lifetime of  $M_z^{\rm tot
}$  at fixed $E_a$ are 
larger than those at fixed $Q_0$ 
by factors  $\ep_{\rm eff}$ and $\ep_{\rm eff}/\ef$, respectively.  
The nonlocal correlations in the surface  
charge density fluctuations 
have  also been found, which are related 
to  the bulk polarization fluctuations in Eq.(133). 
The frequency-dependent dielectric constant 
in small systems has been given in Eqs.(153) and (154).

We comment on future problems. 
(1) We should extend the present theory to 
study the dielectric response in  electrolytes\cite{Leve} 
and elastic dielectrics\cite{Eri}.  
(2)   The nonlinear dielectric 
response\cite{Ful3,Booth,Edholm,Andel1,OnukiD} 
should further be   studied. 
In polar mixtures, the component with 
 a larger  dipole moment tends to accumulate 
in regions of higher  electric field\cite{Andel1,OnukiD}.  
(3) In this paper, the dipole 
density is a constant, but it can be enriched  
around charged objects (electrostriction)\cite{Landau-e,OnukiD}. 
This effect   is   enhanced in supercritical polar  fluids.  
It   even induces  nucleation of liquid droplets around 
 ions  in  metastable vapors\cite{Kita,Leve}.  
(4) In  fluid mixtures, the composition can be strongly 
heterogeneous around charged objects (preferential solvation).
(5) The chemical reactions near surfaces are 
of great interest, where the image interaction can be 
relevant\cite{M2}.

\vspace{2mm}
\noindent 
{\bf Data availability}: The data that support the findings of
this study are available within the article.

\vspace{2mm}
\noindent{\bf Appendix A: 
Thin-film model of Stern layers }\\
\setcounter{equation}{0}
\renewcommand{\theequation}{A\arabic{equation}}
Here, we treat a Stern layer  as a thin  film with a 
 low dielectric constant  $\ep_s(\ll \ep)$  in the region 
$0<z<d~(\ll H)$. 
In the film, the surface charge density ${\sigma}_0$ 
at $z=0$ induces a polarization $p_s$ along the $z$ axis, 
where the  electric field is  given by 
$E_s= 4\pi ({\sigma}_0- p_s)$. 
We then propose  the  dielectric free energy density in the film  as       
\be 
f_s = 2\pi  ({\sigma}_0- p_s)^2 + p_s^2/2\chi_s, 
\en 
where $\chi_s=(\ep_s-1)/4\pi$.  
For  given  ${\sigma}_0$,  minimization of $f_s$ 
 with respect to $p_s$ yields  
\be 
p_s/\chi_s= E_s= 4\pi{{\sigma}}_0/\ep_s, ~~f_s=
2\pi {{\sigma}}_0^2/\ep_s.  
\en 
The surface capacitance is  then 
 given by $C_0 =\sigma_0/dE_s= 
 \ep_s/4\pi d$.    If the dielectric constant varies 
smoothly  as 
$\ep=\ep(z)$, $C_0$ is the $z$-integral 
of $E_s(z)$ in the layer 
and $\Phi_a$ is given by\cite{Mon}  
$\Phi_a= 4\pi\sigma_0\int_0^H dz {\ep(z)}^{-1}.$    
The three-dimensional 
electric potential  
can be calculated if use is made of the Fourier transformation 
in the $xy$ plane\cite{Ful3,T1}. 
Note that the charge-free polarization and electric field\cite{Hamann}, 
 $p_{\rm int}$ and $E_{\rm int}(=-4\pi p_{\rm int})$, 
do not contribute to $f_s$. 

The  integral of $f_s$ in  the film $0<z<d$ is given by 
\be 
d\int d{\bi r}_\perp f_s 
=L^2{\bar \sigma}_0^2/2C_0+ \int d{\bi r}_\perp 
({\sigma}^{\rm inh}_0)^2/2C_0, 
\en 
where   $\sigma_0^{\rm inh}=\sigma_0-{\bar\sigma}_0$ 
is the inhomogeneous part of $\sigma_0$ and   
the first term  gives  the surface free energy 
in Eq.(133).  For  $C_H=C_0$, Eq.(19)  gives the surface electric length,  
\be 
\ell_{\rm w}= 2d(\ep/\ep_s-1)\cong 
2d \ep/\ep_s \gg 2d.
\en  

There should   be Coulombic repulsion 
among the surface charges, which is 
screened by the  polarization deviation near the surface. 
Thus, we have an additional surface free energy density,  
written as  $(\sigma_0^{\rm inh})^2/2D_0$ at long wavelengths. 
Then,  the short-range variance $S_0$ of  ${\sigma}^{\rm inh}_0$  
in Eq.(138) is given by   
\be 
S_0 = (1/C_0+1/D_0)^{-1}. 
\en

We also propose a  dynamic equation for $p_s$,  
\be 
\frac{\p}{\p t} p_s= -L_s \frac{\p f_s}{\p p_s}
= -L_s\Big[ \frac{\ep_s}{\chi_s} p_s- 4\pi{{\sigma}}_0\Big],
\en  
where $L_s$ is a kinetic coefficient and 
 the  relaxation time of ${ p}_s$ is $\tau_s= \chi_s/\ep_s L_s$. 
   For  ${\bar{\sigma}}_0\propto e^{i\omega t}$, 
the  lateral averages of ${ p}_s$ and 
${ E}_s$ in the film are given by 
\be
{\bar p}_s={4\pi\chi_s{\bar{\sigma}}_0}/{\ep_s( 1+i\omega\tau_s) },
~~~ {\bar E}_s
 ={4\pi}{\bar\sigma}_0/{\ep_s^*(\omega)}. 
\en 
We define  the frequency-dependent film dielectric constant 
and surface capacitance by   
\be 
\ep_s^*(\omega)= 1+ \frac{\ep_s-1}{1+i\omega\tau_s\ep_s},~~
C^*(\omega)=\frac{1}{4\pi d}\ep_s^*(\omega). 
\en 
The  induced potential drop at $z=0$ 
 is given by $d{\bar E}_s= {\bar\sigma}_0/C^*(\omega)$, 
Thus, we find Eq.(145).

\vspace{2mm}
\noindent{\bf Appendix B: Onsager theory }\\
\setcounter{equation}{0}
\renewcommand{\theequation}{B\arabic{equation}}

Onsager\cite{Onsager}  supposed   
a spherical cavity 
containing  a single  polarizable polar molecule,   
where  its   volume  $4\pi a^3/3$ was  equated to 
  the inverse  density $1/n$. 
He assumed  the Clausius-Mossotti relation (10)  for the   
 induced polarization of the molecule  
 $n^{-1}{\bi p}_2={\alpha_0}{\bi F}$, 
where   $\alpha_0={\bar\alpha}/n$ is the molecular 
polarizability  in Eq.(10) 
and $\bi F$ is an internal electric field 
given by 
\be 
{\bi F}= {\bar c}{\bi E}+ {\bar r}{\bi p}
={\bar c}{\bi E}+ {\bar r}{\bi p}_1+ {\bar r}{\bar\alpha}{\bi F}.
\en  
Here,  $\bar c$ is  the  cavity factor 
and  $\bar r$ is the reaction factor:  
\be 
{\bar c}= {3\ep}/({2\ep+1}), ~~~ 
{\bar r}={ 8\pi (\ep-1)}/[{3(2\ep+1)}], 
\en 
From  his Eqs.(12), (13), and (20)  
$\bi F$ and the directing field ${\bi E}_d$   are expressed 
in terms of $\bar\alpha$, ${\bar c}$, and ${\bar r}$ as 
\be 
{\bi F}={\bi E}_{\rm d}
+({1-{\bar r}{\bar\alpha}})^{-1}{\bar r} {\bi p}_1,~~
{\bi E}_{\rm d}= 
 ({1-{\bar r}{\bar\alpha}})^{-1} {\bar c}  {\bi E}, 
\en 
which give     Eqs.(23) and (24) from  Eq.(B2). 
Our  $n$, $\bar\alpha$, $\ef$, ${\bi p}_1$, ${\bi p}_2$, and 
${\bi E}_d$ correspond to 
 $N$,  $N\alpha$, $n^2$, $N\mu_0{\bi u}$, $N\alpha{\bi F}$, 
and $(\mu^*/\mu_0){\bi E}$ in his paper,  respectively. 

Within Onsager's scheme, we  introduce $\chi_1$ and $\chi_2$  
by   ${\bi p}_1=\chi_1 {\bi E}$ 
and ${\bi p}_2=\chi_2{\bi E}={\bar\alpha} {\bi F}$.
Then,  Eq.(B1)  gives     
\be 
\chi_1+\chi_2=\chi, ~~ 
\chi_2= {\bar\alpha}(1- {\bar r}{\bar\alpha})^{-1}({\bar c}+ {\bar r}\chi_1).
\en 
These   equations are  solved to give 
\be 
 \chi_1= \chi(1- {\bar r}{\bar\alpha})- 
 {\bar\alpha}{\bar c}, ~~ 
\chi_2= {\bar\alpha} (\chi {\bar r}+ {\bar c}),
\en  
which yield  Eq.(13)  from  Eqs.(10)  and (B2).  
We also find 
$\lambda_{11}=4\pi/3-{{\bar r}}/({1-{\bar r}{\bar\alpha}})$ 
in accord with Eq.(26).

Onsager assumed  a microscopic sphere, 
which is allowable to take  account of  the long-range 
dipolar interaction. 
 Kirkwood and  Fr$\ddot{\rm{o}}$hlich 
assumed a mesoscopic sphere to include the short-range interactions. 
We have also assumed a mesoscopic sphere 
in Sec.IIIB,

\vspace{2mm}
\noindent{\bf Appendix C: Previous linear theories  }\\
\setcounter{equation}{0}
\renewcommand{\theequation}{C\arabic{equation}}

Many authors\cite{Marcus,Tomasi,Kim,Lee,Li,Mat1} 
started with the dielectric  free energy  
without the Lorentz term, which is written as    
\be 
{\cal F}_0 =\int_V  d{\bi r}
\Big[ \frac{1}{8\pi} |{\bi E}|^2  + \frac{1}{2\chi_{01}} 
|{\bi p}_1|^2+ \frac{1}{2\chi_{02}}
|{\bi p}_2|^2\Big]. 
\en  
Here,     $({\bi p}_1, {\bi p}_2,{\chi}_{01}, {\chi}_{02})$  
were  written   as $({\bi p}_u, {\bi p}_e, {\alpha}_u, {\alpha}_e)$ 
by Marcus\cite{Marcus} 
and as  $({\bi p}_h, {\bi p}_e, {\chi}_h, {\chi}_e)$  
 by  Lee and Hynes\cite{Lee}.
 In equilibrium at fixed $Q_0$, this  ${\cal F}_0$  gives 
${\bi p}_i= \chi_{0i}  {\bi E}$  with 
\be 
\chi_{01}= (\ep-\ef)/4\pi,~~~~\chi_{02}=(\ef-1)/4\pi,
\en  
which   differ from $\chi_1$ and $\chi_2$  in Eq.(13) 
 not leading  to   Eq.(9). The expressions  in Eq.(C2) were used 
also by Fr$\ddot{\rm{o}}$hlich\cite{Fro}, Lee-Zwanzig\cite{Zw}, 
  and   Hubbard-Onsager\cite{Hubbard}. 
 
Setting      $\delta{\cal F}_0/\delta{\bi p}_2={\bi 0}$  at fixed ${\bi p}_1$ 
and   ${\bi E}_0$ we find    
\be 
{\bi p}_2= {\bi p}_{2\parallel}= \frac{\ef-1}{4\pi}{\bi E}, ~~~
{\bi E}=\frac{1}{\ef}(
 {\bi E}_0 -{4\pi}  {\bi p}_{1\parallel}) .   
\en  
Substitution of the above $\bi E$ into Eq.(73) gives 
$\tau_L/\tau_D=\ef/\ep$\cite{Hubbard}. 
Under these relations we remove ${\bi p}_2$ from  ${\cal F}_0$ as 
 \bea 
&&\hspace{-11mm} {\cal F}_0=\hspace{-1mm}
\int_V\hspace{-1mm} d{\bi r}\Big[\frac{|{\bi E}_0|^2}{8\pi\ef}
 - \frac{{\bi p}_{1\parallel}\cdot {\bi E}_0}{\ef}  + 
\frac{\ep|{\bi p}_{1{\parallel}}|^2}{2\ef\chi_{01}}
 +\frac{|{\bi p}_{1\perp}|^2}{2\chi_{01}}\Big] . 
\ena  
Compare Eqs.(C3) and (C4)  with  
 Eqs.(63)-(67).

From Eq.(C1) we calculate 
the  correlations (33) as\cite{Lee,OnukiLong} 
\be
{\hat G}^{ij}_{\alpha\beta}({\bi q})
= \chi_{0i}\delta_{ij}\delta_{\alpha\beta}
 -({4\pi} \chi_{0i}\chi_{0j})/{\ep})
 {\hat q}_\alpha{\hat q}_\beta,
\en 
which differ from those in Eq.(38) but 
give ${\hat G}_{\alpha\beta}({\bi q})$ 
in  the form of Eq.(36). In particular, for $i=j=1$,  we have  
\be 
{\hat G}^{11}_{\alpha\beta}({\bi q})= 
[ \delta_{\alpha\beta}-{\hat q}_\alpha{\hat q}_\beta 
+ (\ef/\ep){\hat q}_\alpha{\hat q}_\beta ]
({\ep-\ef})/{4\pi}.
\en  
 The variance of ${\bi M}_1$ in Eq.(45) is given by   
\be 
\av{|{\bi M}_1|^2}/3vk_BT={(\ep-\ef)(2+\ef/\ep)}/{12\pi},
\en  
which is larger than in Eq.(47)  by a factor of  $ (\ef+2)^2/9$. 
Fr$\ddot{\rm{o}}$hlich\cite{Fro} 
derived Eq.(C7) as remarked  in Sec.IIIB. 

  Dinpajooh {\it et al.}\cite{Mat1} expressed   
${\hat G}_{\alpha\beta}^{ij}({\bi q})$  
 differently  from   those in  Eqs.(38) and (C5). 
 For example, they  found    $\chi_{\perp}^{11}= (\ep/\ef-1)/4\pi$ 
and $\chi_{\parallel}^{11}= (1-1/\ep)/4\pi$ 
(written as $\chi_{nn}^{T}$ and  $\chi_{nn}^{L}$ in their paper). 
Their  theory does not  yield  the KF equation  for $\ef\neq 1$.   

\vspace{2mm}
\noindent{\bf Appendix D: Fourier-Laplace 
transforms of relaxation  functions in  linear 
dynamics }\\
\setcounter{equation}{0}
\renewcommand{\theequation}{D\arabic{equation}}
We show how Eqs.(97) and (98) 
 follow in our  continuum theory. 
For the longitudinal parts of ${\bi p}_i$ and ${\bi p}_j$ we  
  consider the  time-correlation functions in the bulk:  
\be  
 G^{ij}_\parallel(t) = 
 \int \hspace{-1mm} d{\bi r}
\av{{\bi p}_{i\parallel}({\bi r},t)\cdot 
{{\bi p}_{j\parallel} ({\bi 0},0)}}/k_BT,
\en 
where    $ G^{ij}_{\parallel}(0) = 
\chi_\parallel^{ij}$ in  Eq.(38) and 
   Eqs.(70) and (71) give    
\be 
\frac{\p}{\p t}  G^{ij}_{\parallel}(t) =-{\sum}_\ell 
{\cal L}_{i\ell}  G^{\ell j}_{\parallel}(t).
\en   
Here,   ${\cal L}_{ij}= L_i (a_{ij}+ 4\pi )$. 
Then, the  FL transform of 
  $G^{ij}_{\parallel}(t) $ is written  as 
${\sum}_\ell U_{i\ell}^*(\omega) \chi_\parallel^{\ell j}$, 
where  $ U_{ij}^*(\omega)$ 
is the inverse of the matrix    
$ i\omega\delta_{ij}+ {\cal L}_{ij}$. 
Here,  ${\sum}_\ell {\cal L}_{i\ell}\chi_\parallel^{\ell j}
=L_i\delta_{ij}$, since  $ \chi_\parallel^{ij}$ is  the inverse  of the matrix 
$ a_{ij}+ 4\pi$. Thus, 
\be 
i\omega \int_0^\infty dt e^{-i\omega t} G^{ij}_{\parallel}(t) =
 \chi_\parallel^{i j}-  U_{ij}^*(\omega)L_j .
\en 
Here,   $p_i^*/E_0^*= {\sum}_j  U_{ij}^*(\omega)L_j$ 
from Eqs.(70),   (71), and (84). 
Thus, $\alpha_i^* (\omega)$ are the 
FL transforms of $-d  G_{i\parallel}(t)/d t$.  
In the same manner,   $\ep_i^* (\omega)$ are the 
FL transforms of $-d G_{i\perp}(t)/dt$.  
The results in Sec.VII follow for  
$L_2\to \infty$.


\begin{thebibliography}{0}
\bibitem{Debye} P.J.W. Debye, {\it Polar Molecules}  
(Chemical Catalog, New York, 1929).
\bibitem{Onsager}
L. Onsager, Electric moments of molecules in liquids, 
J. Am. Chem. Soc. {\bf 58}, 1486 (1936).

\bibitem{Fro}
H. Fr$\ddot{\rm{o}}$hlich, {\it Theory of dielectrics} (Oxford University Press, Oxford, 1949).
\bibitem{Bott} C. J. F. B$\ddot{\rm o}$ttcher, {\it 
Theory of Electric Polarization } (Elsevier, Amsterdam,
1973), Vol. 1.
\bibitem{Kirk}
J. G. Kirkwood, The dielectric polarization of polar liquids,  
J. Chem. Phys. {\bf 7}, 911 (1939).


\bibitem{Felder2} B. U. Felderhof, Fluctuations of partial polarizations 
in dielectrics, J. Phys. C: Solid State Phys. {\bf 12}, 2423 (1979).

\bibitem{Marcus} 
R. A. Marcus, 
Electrostatic free energy and other properties of states 
having nonequilibrium @olarization. I, 
 J. Chem. Phys. J. Chem. Phys. {\bf 24}, 979-989 (1956). 

\bibitem{Lee} 
S. Lee and J. T. Hynes, Solution reaction path Hamiltonian for 
reactions in polar solvents. I. Formulation, 
 J. Chem. Phys. {\bf 88}, 6853-6862 (1988).

\bibitem{Kim} H. J. Kim and  J. T. Hynes, 
Equilibrium and nonequilibrium solvation and solute 
electronic structure. I. Formulation, 
J. Chem. Phys. {\bf 93}, 5194-5210 (1990). 



\bibitem{Tomasi} M.A. Aguilar,  F.J. Olivares del Valle, and  J. Tomasi, 
Nonequilibrium solvation: An ab initio quantum-mechanical 
method in the continuum cavity model approximation, 
J. Chem. Phys.{\bf 98}, 7375-7384 (1993). 

\bibitem{Li} Xiang-Yuan Li, 
 An overview of continuum models for nonequilibrium
solvation: Popular theories and new challenge, 
Int. J. Quantum Chem.{\bf  115}, 700-721 (2015). 

\bibitem{Mat1} M. Dinpajooh, 
 M. D. Newton, and  D. V. Matyushov, 
Free energy functionals for polarization fluctuations: Pekar 
factor revisited,  J. Chem. Phys.{\bf 146}, 064504 (2017). 

\bibitem{OnukiLong} 
A. Onuki, Long-range correlations of polarization 
and number densities in dilute electrolytes,  
 J. Chem. Phys.{\bf 153},  234501 (2020). 








\bibitem{Cole} R. H. Cole, 
Correlation function theory of dielectric relaxation, 
J. Chem. Phys.{\bf 42}, 637-643 (1965). 

\bibitem{Gla} S. H. Glarum, Dielectric relaxation of polar 
liquids, J. Chem. Phys.{\bf 33}, 1371-1275 (1960).


\bibitem{Mason} E. Fatuzzo and P. R. Mason, 
A calculation of the complex dielectric constant of a 
polar liquid by the librating molecule method, 
Proc. Phys. Soc. (London) {\bf 90}, 729-740 (1967). 

\bibitem{Zw} Tsu-Wei  Nee and R. Zwanzig, 
Dielectric relaxation in polar liquids, 
J. Chem. Phys.{\bf 52}, 6353-6363 (1970).


\bibitem{Wer}
M. S. Wertheim, Exact solution of the mean spherical model 
for fluids of hard spheres with permanent electric dipole 
moments, J. Chem. Phys.{\bf 55}, 4291-4298 (1971).




\bibitem{Ni} G. Nienhuis and J, M. Deutch, 
Structure of dielectric fluids.I. The two-particle distribution 
function of polar fluids,  J. Chem. Phys.{\bf 55}, 4213-4229 (1971).

\bibitem{Wi} G. Williams, The use of the dipole correlation function 
in dielectric relaxation, Chem. Rev.{\bf 72}, 55-69 (1972). 

\bibitem{Ful2} R. L. Fulton, On the theory of dielectric relaxation,
 Mol. Phys.{\bf 29}, 405-413 (1975). 

\bibitem{Deutch} 
  D. E. Sullivan and J. M. Deutch, Molecular theory 
of dielectric relaxation,  J. Chem. Phys. {\bf 62}, 2130-2138 (1975).

\bibitem{Neu}  M. Neumann, 
 Dipole moment fluctuation formulas in computer 
simulations of polar systems,  Mol. Phys.{\bf 50}, 841-858 (1983); 
M. Neumann,  G. Steinhauser, and G. S. Pawley, 
Consistent calculation of the static and frequency-dependent 
dielectric constant in computer simulation, 
Mol. Phys.{\bf 52}, 97-113 (1984). 
 

\bibitem{Mukamel} R. F. Loring and S. Mukamel, Molecular theory of 
solvation and dielectric response in polar fluids, 
J. Chem. Phys. {\bf 87}, 1272-1283 (1987).




\bibitem{Mad} P. Madden and D. Kivelson, 
Dielectric friction and molecular reorientation, 
J. Phys. Chem.{\bf 86}, 4244-4256 (1982); A  consistent 
treatment of dielectric phenomena, Adv. Chem, Phys. 
LVI, 467-566 (1984). 



\bibitem{Bagchi} 
B. Bagchi and A. Chandra, 
Polarization relaxation, dielectric dispersion, and
solvation dynamics in dense dipolar liquid,
 J. Chem. Phys.{\bf 90}, 7338-7345 (1989); 
 A. Chandra and B. Bagchi, 
Microscopic free energy functional for polarization 
fluctuations: Generalization of Marcus-Felderhof expression,   
 J. Chem. Phys.{\bf 94}, 2258-2261 (1991).  




  

\bibitem{Elton} D. C. Elton, The origin of the Debye relaxation 
in liquid water and fitting the high-frequency excess response, 
Phys. Chem. Chem. Phys.{\bf 19}, 18739-18749 (2017). 







\bibitem{Berne} B. J. Berne, A self-consistent theory of rotational diffusion, 
 J. Chem. Phys.{\bf 62}, 1154-1160 (1974).


\bibitem{Hubbard} 
J. Hubbard and L. Onsager, Dielectric dispersion and dielectric 
friction in electrolyte solutions.I., 
 J. Chem. Phys.{\bf 67}, 4850-4856 (1977); 
J. Hubbard,  Dielectric dispersion and dielectric 
friction in electrolyte solutions.II., 
J. Chem. Phys.{\bf 68}, 1649-1664 (1978).

\bibitem{Fleming} M.  Maroncelli,  J. Macinnis, G. R. Fleming, 
Polar solvent dynamics and electron-transfer reactions, 
Science {\bf 243}, 1674 (1989). 

\bibitem{Kubo} R. Kubo, Statistical mechanical theory of irreversible 
processes. I, J. Phys. Soc. Jpn., {\bf 12}, 570-586 (1957). 
\bibitem{Kubo1} R. Kubo, Some comments on the dynamical dielectric 
constants, in {\it Cooperative phenomena}, eds. H. Haken and M. Wagner, 
Springer-Verlag, 140-146 (1973). 






\bibitem{Hau} 
J. Hautman, J. W. Halley, and Y. J. Rhee, 
Molecular dynamics simulation 
of water between two ideal classical metal walls, 
 J. Chem. Phys.{\bf 91}, 467-472 (1989).

\bibitem{P1} 
J. W. Perram and M. A. Ratner, Simulations at conducting interfaces: 
Boundary conditions for electrodes 
and  electrolytes,  J. Chem. Phys.{\bf 104}, 5174 (1996).

\bibitem{P3} 
I.-C. Yeh and M. L. Berkowitz, 
Dielectric constant of water at high electric fields: 
Molecular dynamics study, 
J. Chem. Phys.{\bf 110}, 7935 (1999).

\bibitem{P4} 
P. S. Crozier, R. L. Rowley, and D. Henderson, 
Molecular dynamics calculations of the electrochemical 
properties of electrolyte between charged electrodes, 
J. Chem. Phys.{\bf 113}, 9202 (2000).
\bibitem{P5} 
M. K. Petersen, R. Kumar, H. S. White, and G. A. Voth, 
A computationally efficient treatment of polarizable electrochemical 
cells held at a constant potential, 
J. Phys. Chem. C {\bf 116}, 4903 (2012).
\bibitem{P6} 
 S. K. Reed, O. J. Lanning, and  P. A. Madden,
Electrochemical interface between an ionic liquid and a model 
metallic electrode, 
J. Chem. Phys.{\bf 126}, 084704 (2007); 
A. P. Willard, S. K. Reed, P. A. Madden, and D. 
Chandler, Water at an electrochemical interface-a simulation study, 
Faraday Discuss.{\bf 141}, 423-441 (2009).

\bibitem{Limmer}
D. T. Limmer, C. Merlet, M. Salanne, D. Chandler, P. A. Madden, 
R. van Roij, and B. Rotenberg, 
Charge fluctuatins in nanoscale capacitors, 
Phys. Rev. Lett. {\bf 111}, 106102 (2013).
\bibitem{Wang} Z. Wang, Y. Yang, D. L. Olmsted, M.  Asta, B. B. Laird,  
Evaluation of the constant potential method in simulating electric
double-layer capacitors. J. Chem. Phys. 
J. Chem. Phys.{\bf 141}, 184102  (2014). 

 

\bibitem{T1}
K. Takae and A. Onuki, 
Applying electric field to charged and polar particles between 
metallic plates: Extension of the Ewald method,  
 J. Chem. Phys. {\bf 139}, 124108 (2013).
\bibitem{T2}
K. Takae and A. Onuki,  
Molecular dynamics simulation of water between metal walls under 
an electric field: Dielectric response and dynamics after 
field reversal, 
 J. Phys. Chem. B {\bf 119}, 9377 (2015).

\bibitem{Takae}
K. Takae and A. Onuki, Fluctuations of local electric field and 
dipole moments in water between metal walls, 
 J. Chem. Phys. {\bf 143}, 154503 (2015). 

\bibitem{La} J. B. Haskins and  J. W. Lawson, 
Evaluation of molecular dynamics simulation methods for
ionic liquid electric double layers,  
J. Chem. Phys.{\bf 144}, 184707 (2016). 


\bibitem{Le} A. P. dos Santos, M.  Girotto, and Y.  Levin, 
Simulations of Coulomb systems confined by polarizable surfaces using 
periodic Green functions, 
J. Chem. Phys.{\bf 147}, 184105 (2017). 


\bibitem{Sprik1}  C. Zhang, T. Sayer, J. Hutter, and M. Sprik, 
Modelling electrochemical systems with finite field molecular dynamics, 
J. Phys. Energy, {\bf 2}, 032005 (2020); 
T. Dufils, M.  Sprik, and M.  Salanne, 
 Computational amperometry of nanoscale capacitors in molecular simulations,  
J. Phys. Chem. Lett. {\bf 12}, 4357-4361 (2021).


\bibitem{P7} 
S. R. Tee and D. J. Searles, 
Constant potential and constrained charge ensembles for
simulations of conductive electrodes, 
 J. Chem. Theory Comput. {\bf 19}, 2758-2768 (2023). 



\bibitem{Sato} K. Takahashi, H. Nakano, 
and H.  Sato,  Unified polarizable electrode 
models for open and closed circuits: Revisiting 
the effects of electrode polarization and different circuit 
conditions on electrode-electrolyte interfaces. 
J. Chem. Phys. {\bf 157}, 014111 (2022). 

\bibitem{S1} Y.  Tanaka, H.  Sato, and H.  Nakano, 
Computational dielectric spectroscopy on solid-solution
interface by time-dependent voltage applied molecular
dynamics simulation, J. Chem. Phys.{\bf 160}, 144103 (2024)

\bibitem{Holm} 
A.  Reinauer,  S.  Kondrat, and C.  Holm, 
Electrolytes in conducting nanopores: Revisiting constant
charge and constant potential simulations, 
J. Chem. Phys.{\bf 161}, 104101 (2024)


\bibitem{Sprik}
C. Zhang, J.  Hutter, and M. Sprik, 
Computing the Kirkwood g-factor by combining constant Maxwell
electric field and electric displacement simulations: Application to 
the dielectric constant of liquid water, 
J. Phys. Chem. Lett.{\bf  7}, 2696-2701 (2016). 
\bibitem{Cox} 
S. J. Cox and M. Sprik, Finite field formalism for bulk 
electrolyte solutions,  J. Chem. Phys. {\bf 151}, 0644506 (2019).

\bibitem{Cai} J. M. Caillol, Asymptotic behavior of the pair-correlation 
function of a polar fluid, J. Chem. Phys.{\bf 96}, 7039-7053 (1992).


\bibitem{Lebe} J. M. Caillol, D. Levesque, and J. J. Weis, 
Electrical properties of polarizable ionic 
solutions. I. Theoretical aspects, 
 J. Chem. Phys.{\bf 91}, 5544-5554 (1989); 
Electrical properties of polarizable ionic 
solutions. II. Computer simulation results, 
{\bf 91}, 5555-5566 (1989). 



\bibitem{Leeuw} S. W. de Leeuw, J. W. Perram, and E. R. Smith, 
Computer simulation of the static dielectric 
constant of systems with permanent dipole moments, 
Ann. Rev. Phys. Chem. {\bf 37}, 245-270 (1986).


\bibitem{Ku} P. G.  Kusalik, Computer simulation results for the dielectric 
properties of a highly polar fluid, 
J. Chem. Phys.{\bf 93}, 3520-3535 (1990). 



\bibitem{Lada} M. S. Skaf, T. Fonseca, 
 and  B. M. Ladanyi, Wave  vector dependent 
 dielectric relaxation in hydrogen-bonding liquids: 
A molecular dynamics study of methanol, 
J. Chem. Phys.{\bf 98}, 8929-8945 (1993). 
\bibitem{Lada1}  B. M. Ladanyi and M. S. Skaf, 
Wave vector-dependent  
 dielectric relaxation of methanol-water mixtures, 
J. Phys. Chem.{\bf  100}, 1368-1380 (1996). 

\bibitem{Bopp} P. A. Bopp, A. A. Kornyshev, and G. Sutmann, 
Frequency and wave-vector dependent dielectric function of water:
Collective modes and relaxation spectra, 
J. Chem. Phys.  {\bf 109},  1939-1958 (1998)


\bibitem{Lebo} J. L. Lebowitz and  J. K.  Percus, 
 Long-range correlations in a closed system 
with application to nonuniform fluids, 
 Phys. Rev. {\bf 124} 1673-1691 (1961); 
J. L. Lebowitz, J. K.  Percus, and L. Verlet,  
Ensemble dependence of fluctuations with application 
to machine computations.  Phys. Rev.  {\bf 153}, 250-254 (1967).

\bibitem{OnukiP} A. Onuki, 
Extension of Kirkwood-Buff theory: Partial enthalpies,
fluctuations of energy density, temperature, and pressure,
and solute-induced effects in a mixture solvent, 
J. Stat.  Phys. {\bf 191}:68 (2024). 



\bibitem{Hamann} 
C. H. Hamann, A. Hamnett, and W. Vielstich, {\it Electrochemistry}  
(Wiley VCH, 2007).

\bibitem{Behrens} 
S. H. Behrens and M. Borkovec, Electrostatic interaction of colloidal 
surfaces with variable charges, 
J. Phys. Chem. B {\bf 103}, 2918 (1999); 
S. H. Behrens and D. G. Grier, The charge of glass 
and silica surfaces, J. Chem. Phys.{\bf 115}, 6716 (2001).


 \bibitem{Sakuma} H. Itoh and  H. Sakuma, 
Dielectric constant of water as a function of separation in a 
 slab geometry: A molecular dynamics study, 
J. Chem. Phys.{\bf 142}, 184703 (2015)


\bibitem{Maty1}  D. V. Matyushov, 
Dielectric susceptibility in the interface, 
J. Phys. Chem. B {\bf 125}, 8282-8293 (2021). 

\bibitem{Cox1} S. J. Cox  and P. L. Geissler, 
Dielectric response of thin water films:
A thermodynamic perspective, 
Chem. Sci.{\bf 13}, 9102-9111 (2022). 


\bibitem{Laage} 
J.-F.  Olivieri, J. T Hynes, and D. Laage,
 Confined water dielectric constant reduction 
is due to the surrounding low dielectric media 
and not to interfacial molecular ordering. J. Phys. 
 Chem. Lett.{\bf 12}, 4319-4326 (2021). 




\bibitem{Geim} 
L. Fumagalli, A. Esfandiar, 
R. Fabregas, S. Hu1, P. Ares1, A. Janardanan,
Q. Yang, B. Radha, T. Taniguchi, K. Watanabe, G. Gomila,
K. S. Novoselov, and A. K. Geim, 
Anomalously low dielectric constant of confined water,
Science {\bf 360}, 1339-1342 (2018). 



\bibitem{Mon} 
S Mondal and B Bagchi, Dielectric properties in nanoconfined water, 
J. Chem. Phys.{\bf 161},  220901 (2024).

\bibitem{Landau-e} L. D.  Landau and E. M. Lifshitz, 
{\it Electrodynamics of 
Continuous Media} (Pergamon, New York, 1984).

\bibitem{Ma} D. W. R. Gruen and S. Mar$\check{\rm c}$elja, 
Spatially varying polarization in water, 
J. Chem. Faraday Trans. {\bf 2}, 225-242 (1983). 
\bibitem{Ko} A. A. Kornyshev, S. Leikin, and G. Sutmann, 
Overscreening" in a polar fluid as a result of coupling between 
polarization and density fluctuattions, 
 Electrochim. Acta {\bf 42}, 849-865 (1997).


\bibitem{Maggs}  A. C. Maggs and R. Everaers,  Simulating 
nanoscale dielectric response, 
 Phys. Rev. Lett. {\bf 96}, 230603 (2006). 
\bibitem{Onukibook} A. Onuki, 
{\it Phase Transition Dynamics} 
(Cambridge University Press, Cambridge, 2002).

\bibitem{Eri}  J. L. Ericksen, 
 Theory of elastic dielectrics revisited, 
Arch. Rational Mech. Anal. {\bf 183},  299-313(2007). 



\bibitem{OnukiD} 
A. Onuki, Ions and dipoles in electric field: 
nonlinear polarization and field-dependent 
chemical reaction, Eur. Phys. J. E, {\bf  47}:3 (2024). 

\bibitem{Ben} D. Ben-Yaakov,  D. Andelman, 
D. Harries, and R. Podgornik, 
Ions in mixed dielectric solvents: 
density profiles and osmotic pressure between charged 
interfaces, 
J. Phys. Chem. B {\bf 113}, 6001-6011 (2009).


\bibitem{Booth} F. Booth, 
The dielectric constant of water and the saturation effect, 
J. Chem. Phys.{\bf 19}, I391-394 (1951). 


\bibitem{Edholm}  
 L. Sandberg, R. Casemyr, and O. Edholm, 
Calculated hydration free energies of small organic 
molecules using a nonlinear
dielectric continuum Model, 
J. Phys. Chem. B, {\bf 106}, 7889-7897 (2002).
\bibitem{Andel1}
A. Abrashikin, D. Andelman, and H. Orland, Dipolar
Poisson-Boltzmann equations: Ions and dipoles close to
charge interfaces. Phys. Rev. Lett.{\bf 99}, 077801 (2007).



\bibitem{Ful3} R. L. Fulton, 
 Linear and nonlinear dielectric theory for a slab: 
The connections between the phenomenological coefficients 
and the susceptibilities, J. Chem. Phys. {\bf 145}, 084105 (2016). 


\bibitem{Hansen}
V. Ballengger and J.-P. Hansen, 
Local dielectric permittivity near an interface, Europhys. Lett. 
{\bf 63}, 381-387 (2003);  
Dielectric permittivity profiles 
of confined fluids, J. Chem. Phys. {\bf 122}, 114711 (2005).  

\bibitem{Ful4} R. L. Fulton, 
 Polarization fluctuations in regions 
bounded by equi-potentials, 
 Physica A {\bf 97}, 189-194 (1979). 


\bibitem{M2} 
R. A. Marcus, 
 Reorganization free energy for electron 
transfers at liquid-liquid and dielectric 
semiconductor-liquid interfaces, 
J. Phys. Chem.{\bf 94}, 1050-1055 (1990). 
\bibitem{N1} 
Yi-P.  Liu and M. D. Newton, 
  Reorganization energy for electron transfer at film-modified 
electrode surfaces: A dielectric continuum model, 
J. Phys. Chem. {\bf 98}, 7162-7169 (1994). 



\bibitem{Ful1} 
R. L. Fulton, 
Susceptibilities, polarization fluctuations, and the dielectric 
constant in polar media, 
 J. Chem. Phys.{\bf 63}, 77-82 (1975).

\bibitem{Felder1}
B. U. Felderhof, Fluctuations of polarization and magnetization 
in dielectric and magnetic media, 
J. Chem. Phys. {\bf 67}, 493-500 (1977).

\bibitem{Landau-s} L.D. Landau and E.M. Lifshitz, 
{\it Statistical Physics} (Pergamon, New York, 1964).  



\bibitem{Boyd} R. H. Boyd, Extension of Stokes' law for ionic motion 
to include the effect of dielectric relaxation, 
J. Chem. Phys. {\bf 35}, 1281-1283 (1961).

\bibitem{Zw1}
R. Zwanzig, Dielectric friction on a moving ion, 
J. Chem. Phys. {\bf 38}, 1603-1605 (1963); 
Dielectric friction on a rotating dipole, 
J. Chem. Phys. {\bf 38}, 1605-1606 (1963).

\bibitem{Wo} J. B. Hubbard and 
P. G. Wolynes, Dielectric friction 
and molecular reorientation, 
 J. Chem. Phys.{\bf 69}, 998-1006 (1978). 
. 


\bibitem{Felder3}
B. U. Felderhof,
Dielectric friction on a polar molecule 
rotating i a fluid,  Mol. Phys. {\bf 48}, 1269-1281 (1983). 
 

\bibitem{Zwan} G. van der Zwan and J. T. Hynes, Polarization 
diffusion and dielectric friction, Physica A {\bf 121}, 227-252 (1983). 
 
\bibitem{Mat2} D. V. Matyushov, 
 Electrophoretic mobility of nanoparticles in water 
J. Phys. Chem. B, {\bf 128}, 2930-2938 (2024).


\bibitem{Wolynes} 
D. F. Calel and P. G. Wolynes, Smoluchowski-Vlasov theory of charge 
solvation dynamics,  
 J. Chem. Phys.{\bf 78}, 4145-4153 (1983). 





\bibitem{Roy} S. Roy and B. Bagchi, Solvation dynamics 
in liquid water. A novel interplay between liberation and diffusive 
motions, J. Chem. Phys.{\bf 99}, 9938-9943 (1993).


\bibitem{Alva} F. Alvarez,  A. Arbe, and  J. Colmenero, 
The Debye's  model for the dielectric relaxation of liquid 
water and the role of cross-dipolar correlations. 
A MD-simulations study, J. Chem. Phys.{\bf 159}, 134505 (2023). 

\bibitem{Feller} H. A. Stern and S. E. Feller, 
Calculation of the dielectric permittivity profile for a 
nonuniform system: Application to a lipid bilayr simulation, 
J. Chem. Phys. {\bf 118}, 3401 (2003).



\bibitem{Netz}
D. J.  Bonthuis, S. Gekle, and R. R. Netz, 
Profile of the Static Permittivity Tensor of Water at Interfaces:
Consequences for Capacitance, Hydration Interaction and Ion
Adsorption,   Langmuir {\bf 28}, 7679 (2012). 


\bibitem{Klapp} V. A. Frolistov ans S. H. L.  Klapp, 
Dielectric response of polar liquids in narrow slit pores, 
J. Chem. Phys. {\bf 126}, 114703 (2007).


\bibitem{Netz1}
S. Gekle and R. Netz, , Anisotropy in the dielectric 
spectrum of hydration water and its relation to water dynamics, 
J. Chem. Phys. {\bf 137}, 104704 (2012).




\bibitem{Sprik2} 
J. I.  Siepmann and M. Sprik, Influence of surface topology 
and electrostatic potential on water$/$electrode systems, 
J. Chem. Phys. {\bf 102}, 511-524 (1995).  




\bibitem{Kita} H. Kitamura and A.  Onuki, 
Ion-induced nucleation in polar one-component fluids, 
J. Chem. Phys.  {\bf  123}, 124513 (2005).

\bibitem{Leve} A. K. Shchekin and T. S. Levedeva, 
Density functional description of size-dependent effects 
at nucleation on neutral and charged nanoparticles, 
J. Chem. Phys.  {\bf  146}, 094702 (2017).

\end{thebibliography}
\end{document}